\documentclass[journal=mamobx,manuscript=article, layout=traditional]{achemso}

\usepackage{amsmath,amssymb}
\usepackage{graphicx}
\usepackage[colorlinks=true,linkcolor=blue,citecolor=blue,urlcolor=blue]{hyperref}

\usepackage{bm}

\usepackage[version=3]{mhchem} 

\usepackage{color}
\usepackage{ulem}
\usepackage{flafter}

\SectionNumbersOn

\newcommand{\onlinecite}[1]{\hspace{-1 ex} \nocite{#1}\citenum{#1}}

\newcommand{\doi}[1]{\href{http://dx.doi.org/#1}{\nolinkurl{#1}}}

\title[Viscosity Overshoot in Biaxial Elongational Flow]{
Viscosity Overshoot in Biaxial Elongational Flow:
Coarse-Grained Molecular Dynamics Simulation
of Ring-Linear Polymer Mixtures
}

\author{T. Murashima}
\email{murasima@cmpt.phys.tohoku.ac.jp}
\phone{+81-22-795-5718}
\fax{+81-22-795-6447}
\affiliation{Department of Physics, Tohoku University, 6-3, Aramaki-aza-Aoba, Aoba-ku, Sendai, 980-8578, Japan}

\author{K. Hagita}
\affiliation{Department of Applied Physics, National Defense Academy, 1-10-20, Hashirimizu, Yokosuka, 239-8686, Japan}

\author{T. Kawakatsu}
\affiliation{Department of Physics, Tohoku University, 6-3, Aramaki-aza-Aoba, Aoba-ku, Sendai, 980-8578, Japan}


\begin{document}

%

\begin{center}
\vspace*{20pt}
\textbf{Graphical TOC Entry}
\vspace*{10pt}

\fbox{
\includegraphics[width=3.25in,height=1.375in]{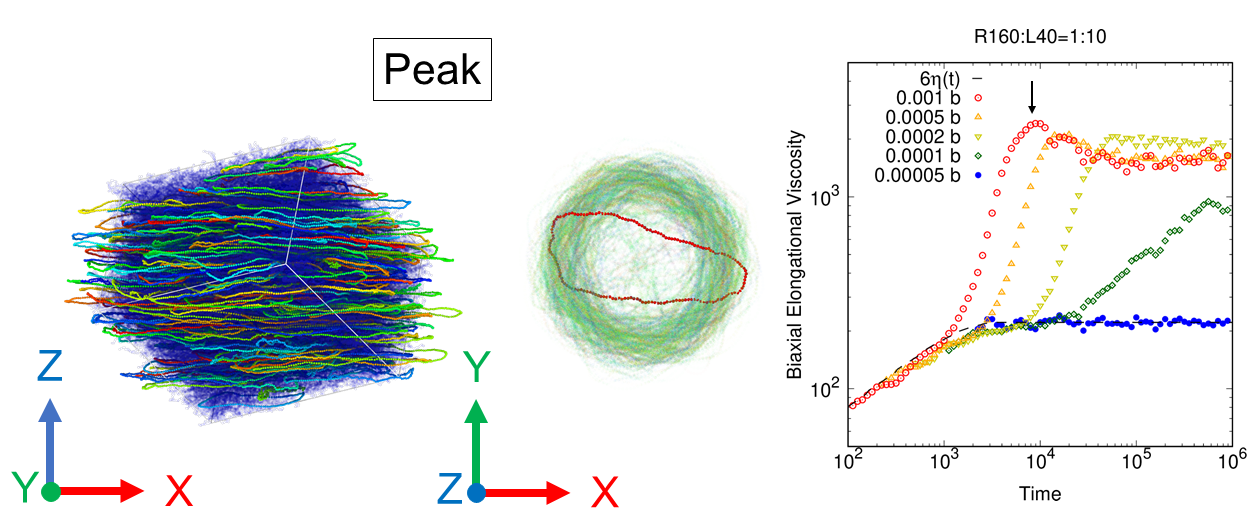}
}

(for Table of Contents use only)
\end{center} 


\clearpage

\begin{abstract}
Viscosity overshoot of entangled polymer melts has been observed under shear flow and uniaxial elongational flow, but has never been observed under biaxial elongational flow.
We confirmed the presence of viscosity overshoot under biaxial elongational flows observed in a mixed system of ring and linear polymers expressed by coarse-grained molecular dynamics simulations. The overshoot was found to be more pronounced in weakly entangled melts. 
Furthermore, the threshold strain rate $\dot{\varepsilon}_{\rm th}$ distinguishing linear and nonlinear behaviors was found to be dependent on the linear chain length as $\dot{\varepsilon}_{\rm th}(N)\sim N^{-1/2}$, which differs from the conventional relationship, $\dot{\varepsilon}_{\rm th}(N) \sim N^{-2}$, expected from the inverse of the Rouse relaxation time. 
We have concluded that the cooperative interactions between rings and linear chains were enhanced under biaxial elongational flow.
\end{abstract}

\section{Introduction}

Entangled polymer melts exhibit nonlinear flow behavior depending on their molecular structures and compositions~\cite{DoiEdwards,Utracki,Macosko,Larson}. It is well known that many entangled polymers exhibit viscosity overshoot under shear flow~\cite{Macosko,Larson}. Branched polymers exhibit viscosity overshoot under uniaxial elongational flow~\cite{RasmussenNielsenBachHassager2005}. However, viscosity overshoot behavior has never been reported 
under biaxial elongational (uniaxial compression) flow.

The origin of viscosity overshoot in shear flow has been intensively investigated~\cite{MasubuchiWatanabe2014,CaoLikhtman2015,WatanabeMatsumiyaInoue2016}. When shear flow is first applied, entanglements among polymers result in a temporary excess orientation of subchains of the polymer. 
In shear flow, the overly oriented subchains are relaxed by the well-known convective constraint release (CCR) mechanism~\cite{Marrucci1996,IannirubertoMarrucci1996}.
The CCR mechanism tells us that the relaxation of orientation is accelerated when the flow rate is faster than the thermal rate.
Viscosity is increased by the excess orientation, and then attenuated by the relaxation of the overly oriented subchains.
The viscosity reaches a steady state when the forced orientation by the outside field and the relaxation of excess orientation are balanced.

Viscosity overshoots are also observed in nonlinear elongational flows of branched polymers~\cite{RasmussenNielsenBachHassager2005}.  The mechanism can be explained using the POM-POM model~\cite{HawkeHuangHassagerRead2015}.
Under uniaxial elongational deformation, the backbone first stretches in the elongational direction and the tension increases. Then, as the arms of the branch begin to be drawn into the backbone tube, the tension relaxes. Once the branched arms are completely inside the backbone tube, the viscosity enters steady state, and the branched polymer behaves similar to a linear polymer. 

For viscosity overshoot to occur under biaxial elongational flow, a mechanism depending on the molecular architecture must exist to maximize tension in a two-dimensional plane parallel to the elongational directions.

Ring polymers are likely to exhibit viscosity overshoot under biaxial elongational deformation due to the maximization of tension in the two-dimensional plane. Recently, ring polymers have attracted much attention because their mechanical response is quite different from that observed in conventional polymer systems. The stress relaxation of ring polymers exhibits a power law decay over long duration due to a phenomenon called ``threading'' where the polymers interpenetrate each other~\cite{HalversonLeeGrestGrosbergKremer2011,LeeKimJung2015}. Ring polymer melts are more easily forming a glassy state at low temperatures~\cite{MichielettoTurner2016}. It has also been reported that when a small amount of ring polymer is added to the linear polymer melt, the relaxation time increases, and the viscosity increases in proportion to the concentration of the ring polymer~\cite{HalversonGrestGrosbergKremer2012,ParisiAhnChangVlassopoulosRubinstein2020}. Under shear flows, hydrodynamic interactions inflate ring polymers in a solvent~\cite{LiebetreuLikos2020} and a linear chain matrix~\cite{WangZhaiChenWangLuAn2019}. The inflation of DNA ring polymers has also been observed in semidilute linear polymer solutions under planar elongational flow~\cite{ZhouHsiaoReganKongMcKennaRobertson-AndersonSchroeder2019}.

In molecular dynamics simulations, it has been difficult to achieve large strains in elongational flows, except for planar elongation, due to the problem of boundary conditions. The Kraynik--Reinelt (KR) boundary conditions~\cite{KraynikReinelt1992}, which has been used for planar elongational deformation~\cite{BaranyaiCummings1999,MatinDaivisTodd2000}, has been extended by Dobson~\cite{Dobson2014} and Hunt~\cite{Hunt2015} to handle uniaxial elongational deformation and biaxial elongational deformation in molecular dynamics simulations. The simulation code was provided by Nicholson and Rutledge~\cite{NicholsonRutledge2016}, which led to intensive studies on extensional flows of polymer melts~\cite{OConnorAlvarezRobbins2018,MurashimaHagitaKawakatsu2018,OConnorHopkinsRobbins2019,OConnorGeRubinsteinGrest2020,Borgeretal2020}. Especially for ring polymer melts, O’Connor et al. found anomalous thickening behaviors owing to “reef knots”~\cite{SubramanianShanbhag2008} or persistent threadings comprising two rings in uniaxial elongational flow~\cite{OConnorGeRubinsteinGrest2020}. Very recently, Borger et al. found a threading-unthreading transition for ring-linear blends in uniaxial elongational flow~\cite{Borgeretal2020}, where the formation of persistent threading was suppressed to form in ring-linear blends. 
Because the study of ring-linear blends in elongational flows is just beginning, new phenomena are expected to be discovered in the other elongational flows.

In this study, we focus on ring-linear blends in biaxial elongational flow. The ring fraction of blends studied here is set to less than 0.1. This fraction is smaller than that employed in a previous work, where uniaxial elongational flow was associated with stress overshoot phenomena~\cite{Borgeretal2020}. 
To exclude the possibility of the persistent threading observed in a previous work~\cite{OConnorGeRubinsteinGrest2020} and to focus on ring-linear interactions in short chain matrices rather than the entangled network~\cite{Borgeretal2020}, we choose the ring fraction to be smaller than the closest packing structure of rings, when a ring polymer is approximately represented as a sphere.
The melt mixtures of ring and linear polymers under biaxial elongational flow were analyzed by coarse-grained molecular dynamics simulations, and the changes in stress and viscosity over time were investigated. 
We observed viscosity overshoot phenomena under biaxial elongational flows in weakly entangled melts.
Our analyses have revealed that the origins lie in the cooperative interactions of ring and linear polymers.

\section{Simulation Methods}

\subsection{Kremer--Grest type bead-spring model under flow field}

Ring and linear polymers were represented by a bead-spring model, namely, the Kremer--Grest model~\cite{KremerGrest1990}. Ring and linear polymers are composed of $N_{\rm R}$ and $N_{\rm L}$ particles connected with $N_{\rm R}$ and $N_{\rm L}-1$ bonds, respectively. The numbers of ring and linear polymers in the system are $M_{\rm R}$ and $M_{\rm L}$, respectively. The polymers are placed in a cubic unit cell $\boldsymbol{L}={\rm diag}(L,L,L)$ with periodic boundary conditions, where $L=V^{1/3}$ and $V$ is the volume of unit cell.
The temperature is controlled by the Langevin thermostat.

The dynamics of particles in equilibrium are governed by Newton’s equation of motion:
\begin{align}
\frac{{\rm d}\boldsymbol{r}}{{\rm d}t}&=\boldsymbol{v},\\
m\frac{{\rm d}\boldsymbol{v}}{{\rm d}t}&=\boldsymbol{F},
\end{align}
where $\boldsymbol{r}$ is the position, $\boldsymbol{v}$ is the velocity, and $m$ is the mass of the particle. The force $\boldsymbol{F}$ is composed of a conservative force, a viscous force, and a random force:
\begin{equation}
    \boldsymbol{F}=-\boldsymbol{\nabla}U -\Gamma \boldsymbol{v} 
    + \sqrt{2\Gamma k_{\rm B} T} \boldsymbol{R}(t),\label{eq:force}
\end{equation}
where $\Gamma$ is the damping coefficient, $k_{\rm B}$ is Boltzmann’s constant, and $T$ is the temperature. The random vector $\boldsymbol{R}(t)$ has a Gaussian probability distribution function with correlation function $\langle R_i(t) R_j(t') \rangle = \delta_{ij} \delta(t-t')$. The potential energy $U(r)$ consists of a Lennard-Jones (LJ) potential $U_{\rm LJ}(r)$ among intramolecular and intermolecular particles and a finite extensible nonlinear elastic (FENE) potential $U_{\rm FENE}(r)$ for particles connected with a bond:
\begin{align}
U(r)&=\sum_{\text{intra + inter}} U_{\rm LJ}(r) + \sum_{\text{intra}} U_{\rm FENE}(r) \\
U_{\rm LJ}(r)&=
\begin{cases}
\displaystyle
4\epsilon \left[ \left(\frac{\sigma}{r} \right)^{12}-\left(\frac{\sigma}{r}\right)^{6} \right]
+\epsilon, & (r<r_{\rm c}) \\
0, & (r \ge r_{\rm c})
\end{cases}
\\
U_{\rm FENE}(r)&=
\begin{cases}
\displaystyle
-\frac{k}{2}R_0^2 \ln 
\left[
1-\left(\frac{r}{R_0} \right)
\right], & (r<R_0) \\
0, & (r\ge R_0)
\end{cases}
\end{align}
where $\epsilon$ is the unit of energy, $\sigma$ is the size of the particle, $r_{\rm c}$ is the cutoff length of the LJ potential, $k$ is the bond coefficient, and $R_0$ is the maximum extent of the FENE bond. To investigate the rheological properties, we compute the stress tensor 
\begin{equation}
\sigma_{\alpha \beta}=\frac{-1}{V}
\left(
\sum_{i=1}^{N_{\rm total}}m v_i^{\alpha} v_i^{\beta} 
+\sum_{j=1}^{N'_{\rm total}}r_j^{\alpha}F_j^{\beta}
\right),
\end{equation}
where $N_{\rm total}$ is the total number of LJ particles in the system, and $N'_{\rm total}$ includes the periodic image particles.

We set the parameters to the conventional ones: $\Gamma=0.5 m/\tau$, $T=1.0\epsilon/k_{\rm B}$, $r_{\rm c}=2^{1/6}\sigma$, $k=30.0\epsilon/\sigma^2$, $R_0=1.5\sigma$, number density $\rho=N_{\rm total}/V=0.85 /\sigma^{3}$, unit of time in LJ system $\tau=\sqrt{m\sigma^2/\epsilon}$, and LJ unit $\sigma=m=\epsilon=k_{\rm B}=1$.

In a flow field $\mathbf{K}=(\boldsymbol{\nabla}\boldsymbol{v})^T$, Newton’s equation of motion is modified to the SLLOD equation~\cite{EvansMorriss}:
\begin{align}
\frac{{\rm d}\boldsymbol{r}}{{\rm d}t}&=\boldsymbol{v}
+\mathbf{K}\cdot\boldsymbol{r}, \label{eq.SLLOD1}\\
\frac{{\rm d}\boldsymbol{v}}{{\rm d}t}&=\frac{\boldsymbol{F}}{m}-\mathbf{K}\cdot\boldsymbol{v}.\label{eq.SLLOD2}
\end{align}
For the biaxial elongational flow with elongational rate $\dot{\varepsilon}$, the flow field $\mathbf{K}$ is represented as $\mathbf{K}={\rm diag}(\dot{\varepsilon}, \dot{\varepsilon}, -2\dot{\varepsilon})$.
To apply the elongational deformation to the system for long duration, we use the generalized Kraynik--Reinelt (gKR) boundary conditions~\cite{Dobson2014,Hunt2015,NicholsonRutledge2016,MurashimaHagitaKawakatsu2018} to prevent the collapse of the simulation box.
The details of the gKR boundary conditions are summarized in Appendix \ref{App:gKR}. 
The code, USER-UEF, developed by Nicholson and Rutledge~\cite{NicholsonRutledge2016} is available for the Nos\'{e}--Hoover thermostat.
We have developed a code, USER-UEFEX, that is suitable for the Langevin thermostat under the gKR boundary conditions~\cite{MurashimaHagitaKawakatsu2018}.

\subsection{Simulation setups}

Here, we consider a ring with 160 beads (called R160) and linear chains with 10, 20, 40, 80, and 160 beads (L10, L20, L40, L80, and L160). We investigate 1:10 blends of ring and linear polymers (R160-L10, R160-L20, R160-L40, R160-L80, and R160-L160). 
The total number of beads in the system $M_{\rm R}N_{\rm R} + M_{\rm L}N_{\rm L}$ is fixed as 675,840, and $M_{\rm R}N_{\rm R}$:$M_{\rm L}N_{\rm L}$ = 1:10 = 61,440:614,400.
The numbers of molecules in the ring-linear blends considered here are then as follows: $M_{\rm R160}=384$, 
$M_{\rm L10}=61,440$, 
$M_{\rm L20}=30,720$,
$M_{\rm L40}=15,360$,
$M_{\rm L80}=7,680$,
and $M_{\rm L160}=3,840$.
The combinations of beads and molecules in ring-linear blends are summarized in Table~\ref{table_number}.

The actual ring fraction is $1/11 \approx$ 0.091, which is smaller than 0.1.
From the radius of gyration $R_{\rm g}$, we can estimate the effective volume of the rings $V_{\rm eff}=\frac{4\pi}{3}R_{\rm g}^3$.
From our recent work~\cite{HagitaMurashima2021}, the radius of gyration $R_{\rm g}$ in ring-linear blends depends on the linear chain length and lies between 4.85 (R160-L160) and 5.10 (R160-L10).
Thus, $V_{\rm eff}$ lies between 478 and 556.
The total effective volume of the rings $V_{\rm eff} M_{\rm R160}$ (between $1.8\times 10^5$ and $2.1 \times 10^5$) is a fourth of the total volume ($8.0\times 10^5$ with density 0.85).
Because the rings are homogeneously dispersed in the ring-linear blends,
we can avoid the ring--ring interpenetration in this fraction in equilibrium state.

To prepare ring-linear blends in equilibrium states, we conducted more than $1.0\times10^9$ equilibration runs or $1.0\times10^7 \tau$, which is much larger than the longest relaxation time of L160, $\tau_{1,\rm L160}\approx 5.5\times 10^4 \tau$ (see Ref.~\onlinecite{MurashimaHagitaKawakatsu2018}). We have confirmed in our recent work~\cite{HagitaMurashima2021} that at least $1.0\times 10^9$ steps are required to achieve an equilibrium distribution of linear chain penetration through rings. For long-duration equilibration, we used Large-scale Atomic/Molecular Massively Parallel Simulator (LAMMPS)~\cite{Plimpton1995} and HOOMD-blue~\cite{AndersonGlaserGlotzer2020}.

To apply a biaxial elongational flow with an elongational flow rate $\dot{\varepsilon}$ to the equilibrated system, we used LAMMPS with USER-UEFEX~\cite{MurashimaHagitaKawakatsu2018}. The biaxial elongational flow rates were selected as $\dot{\varepsilon}=\{1\times10^{-5},2\times10^{-5}, 5\times10^{-5},1\times10^{-4},2\times10^{-4}, 5\times10^{-4},1\times10^{-3} \} \tau^{-1}$,
which includes the inverse of the entanglement time $\tau_{\rm e}\approx 2000 \tau$ (see Ref.~\onlinecite{KremerGrest1990}) and the inverse of the maximum relaxation time $\tau_{1,\rm L160}$ in this study, allowing nonlinear behaviors to be captured. Each elongational flow simulation was conducted on $1.0\times10^8$ steps.

\begin{figure}[htbp]
    \centering
    \includegraphics[width=7in]{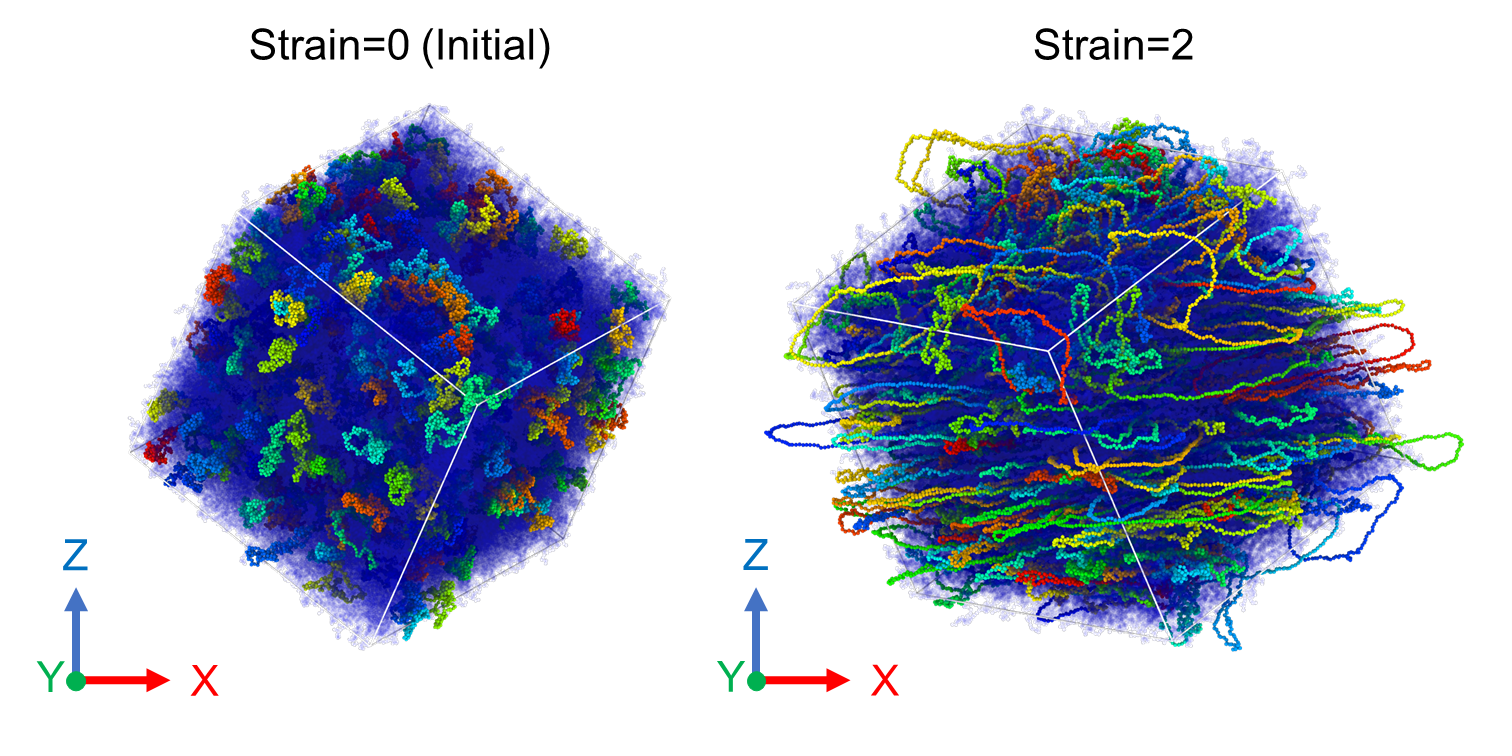}
    \caption{Snapshots of R160-L40 at the initial condition (left) and when the strain is 2 (right) under biaxial elongational flow ($\dot{\varepsilon}=0.001$) with the gKR boundary conditions. Ring polymers are highlighted with colors, and linear polymers are represented with transparent particles.
    We used OVITO~\cite{Alexander2010} for visualization.}
    \label{fig1_snap}
\end{figure}

To facilitate the reader's understanding, typical snapshots of the blend of R160-L40 under biaxial elongational flow ($\dot{\varepsilon}=0.001$) with the gKR boundary conditions are shown in Fig.~\ref{fig1_snap}. Under the gKR boundary conditions, the initial simulation box (the left figure in Fig.~\ref{fig1_snap}) is not placed parallel to the elongational flow directions as mentioned in Appendix \ref{App:gKR}.
Applying the biaxial strain parallel to the $xy$-plane, the ring and linear polymers are expanded parallel to the flow directions as shown in the right figure in Fig.~\ref{fig1_snap}.
This behavior is associated with the hydrodynamic inflation of ring polymers observed in the shear flows~\cite{LiebetreuLikos2020,WangZhaiChenWangLuAn2019} and the planar elongational flow~\cite{ZhouHsiaoReganKongMcKennaRobertson-AndersonSchroeder2019}.

\begin{table*}[htbp]
\centering
\caption{Numbers of beads $N$ and molecules $M$ in ring-linear blends}
\begin{tabular}{crrrrr}
\hline
    & R160-L10 & R160-L20 & R160-L40 & R160-L80 & R160-L160 \\
\hline
$N_{\rm R}$ & 160 & 160 & 160 & 160 & 160  \\
$M_{\rm R}$ & 384 & 384 & 384 & 384 & 384 \\
$N_{\rm L}$ & 10 & 20 & 40 & 80 & 160 \\
$M_{\rm L}$ & 61,440 & 30,720 & 15,360 & 7,680 & 3,840 \\
\hline
\end{tabular}
\label{table_number}
\end{table*}

\clearpage

\section{Results and Discussion}

\subsection{Relaxation modulus (linear viscoelasticity)}

\begin{figure}[htbp]
    \centering
    \includegraphics[width=3.5in]{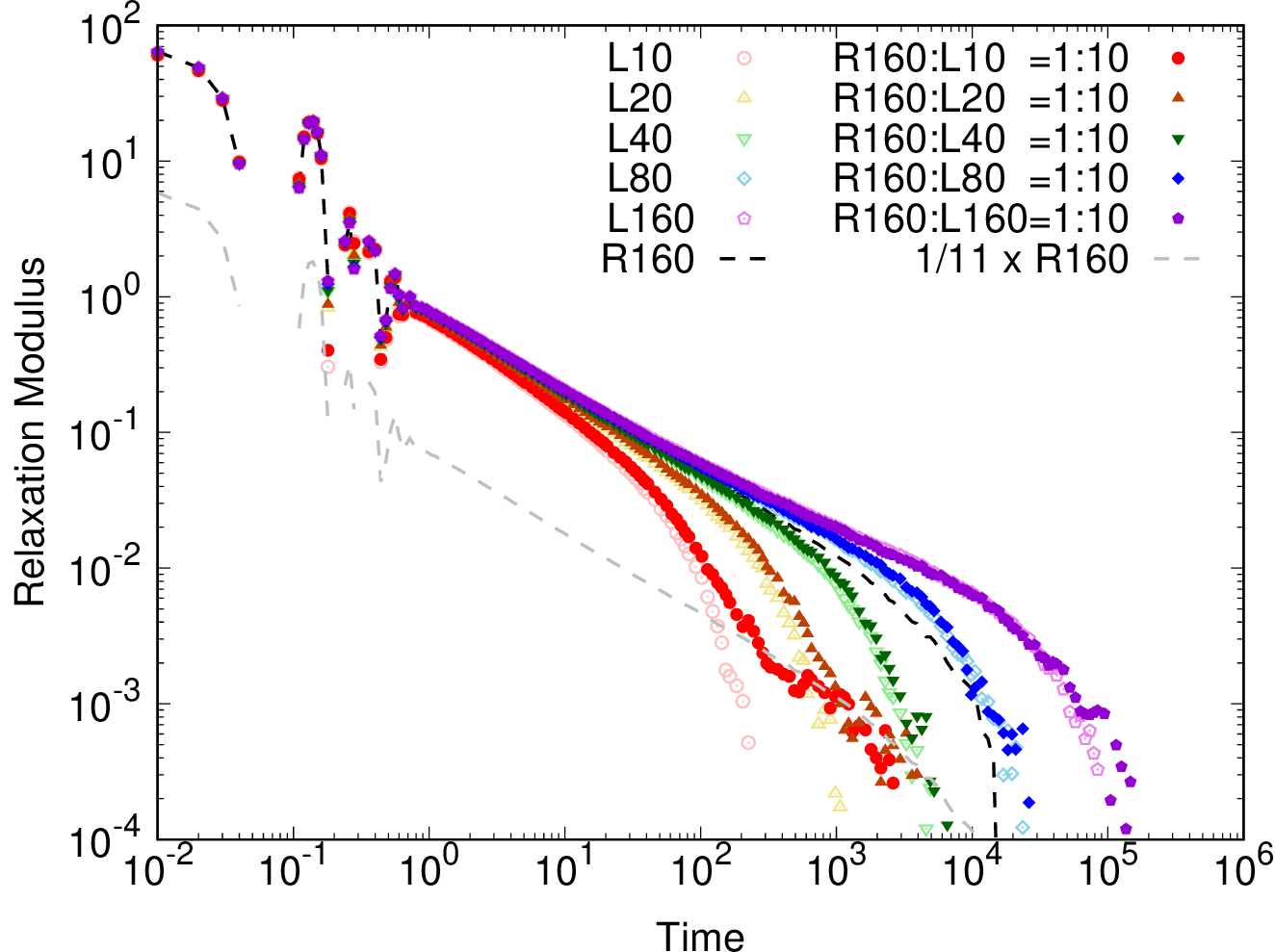}
    \caption{Relaxation modulus of pure linear melts (L10, L20, L40, L80, L160), pure ring melt (R160), and ring-linear blends (R160-L10, R160-L20, R160-L40, R160-L80, R160-L160). The one-eleventh values of R160 were plotted as a guide for the eye.}
    \label{fig2_gt}
\end{figure}

Before discussing the biaxial elongational flows, we would like to summarize the static properties of ring-linear blends. From the Green--Kubo relation on stress autocorrelation functions in equilibrium states, we can obtain the relaxation modulus, 
\begin{equation}
 G(t)=\frac{V}{k_{\rm B}T}\langle  \sigma_{xy} (t) \sigma_{xy} (0) \rangle,   
\end{equation}
where the stress autocorrelation functions are calculated by the multiple-tau correlator method~\cite{RamirezSukumaranVorselaarsLikhtman2010}.
This method is an efficient on-the-fly algorithm, and the stress autocorrelation functions are computed while the equilibrium simulation is run. To obtain the stress autocorrelation functions, we conducted an extra equilibrium simulation with $1.0\times 10^9$ steps.
Figure \ref{fig2_gt} summarizes the relaxation moduli of L10, L20, L40, L80, L160, R160, R160-L10, R160-L20, R160-L40, R160-L80, and R160-L160. One-eleventh of the relaxation modulus of R160 is also plotted in Fig.~\ref{fig2_gt} as a guide for the eye. 
To improve the visibility of graph, the fluctuating data after the terminal relaxation are not displayed in Fig.~\ref{fig2_gt}.
From the relaxation modulus $G(t)$, we can obtain the storage modulus $G'(\omega)$ and the loss modulus $G''(\omega)$.
These moduli, $G'(\omega)$ and $G''(\omega)$, can be useful for experimental testing, and they are presented in Appendix \ref{App:G1G2}.

The relaxation moduli of R160-L10 and R160-L20 clearly show two-step relaxation, whereas those of R160-L40, R160-L80, and R160-L160 do not show it. 
For R160-L10, R160-L20, and R160-L40, their terminal relaxations overlap with the one-eleventh values of R160.
The relaxation moduli of R160-L40, R160-L80, and R160-L160 slightly deviate from those of L40, L80, and L160, respectively. However, the deviation is negligibly small. We found that the ring polymer, R160, introduces slow modes into a linear matrix of L10 and L20, whereas the contributions of rings in L40, L80, and L160 are not apparent within the linear viscoelasticity.

\clearpage

\subsection{Observation of viscosity and stress overshoots under biaxial elongational flows}

\begin{figure*}[htbp]
    \centering
    \includegraphics[width=7in]{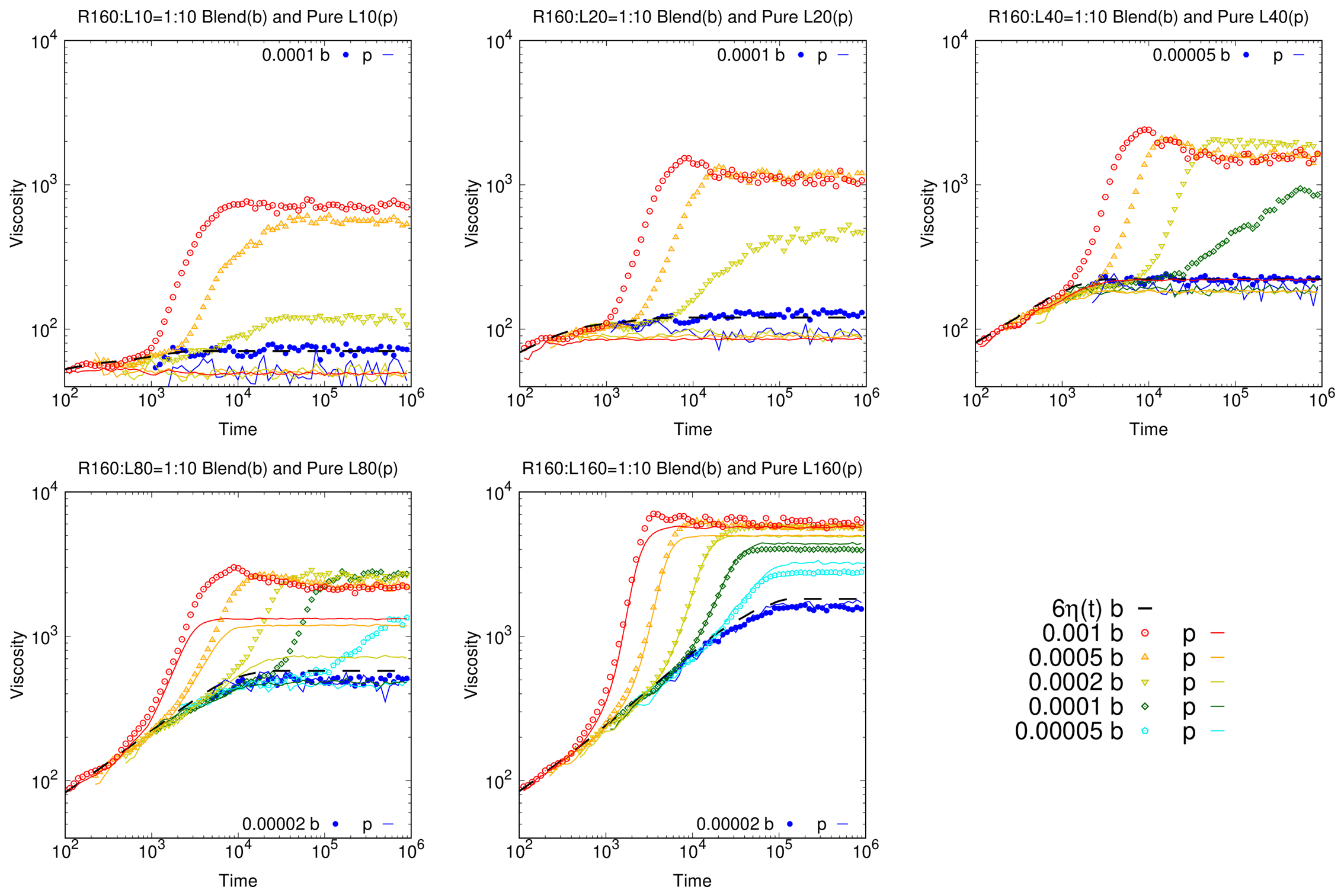}
    \caption{Viscosity growth (viscosity--time) curves under biaxial elongational flows with biaxial elongational rates $\dot{\varepsilon}$. R160-L10, R160-L20, and R160-L40 blends are shown in the top row, and R160-L80 and R160-L160 are in the bottom row. Each color represents the corresponding elogation rate $\dot{\varepsilon}$. Symbols represent data of blends, and lines represent data of pure linear melts (L10, L20, L40, L80, and L160) for comparison among ring-linear blends (b) and pure melts of linear chains (p). The dashed lines represent the linear viscosity growth curve of biaxial elongational flow, $6\eta(t)$.
    Filled circles in each graph represent the threshold strain rate $\dot{\varepsilon}_{\rm th}$. The data in the linear regime $\dot{\varepsilon} < \dot{\varepsilon}_{\rm th}$ for ring-linear blends are not presented.}
    \label{fig3_visc}
\end{figure*}

\begin{figure*}[htbp]
    \centering
    \includegraphics[width=7in]{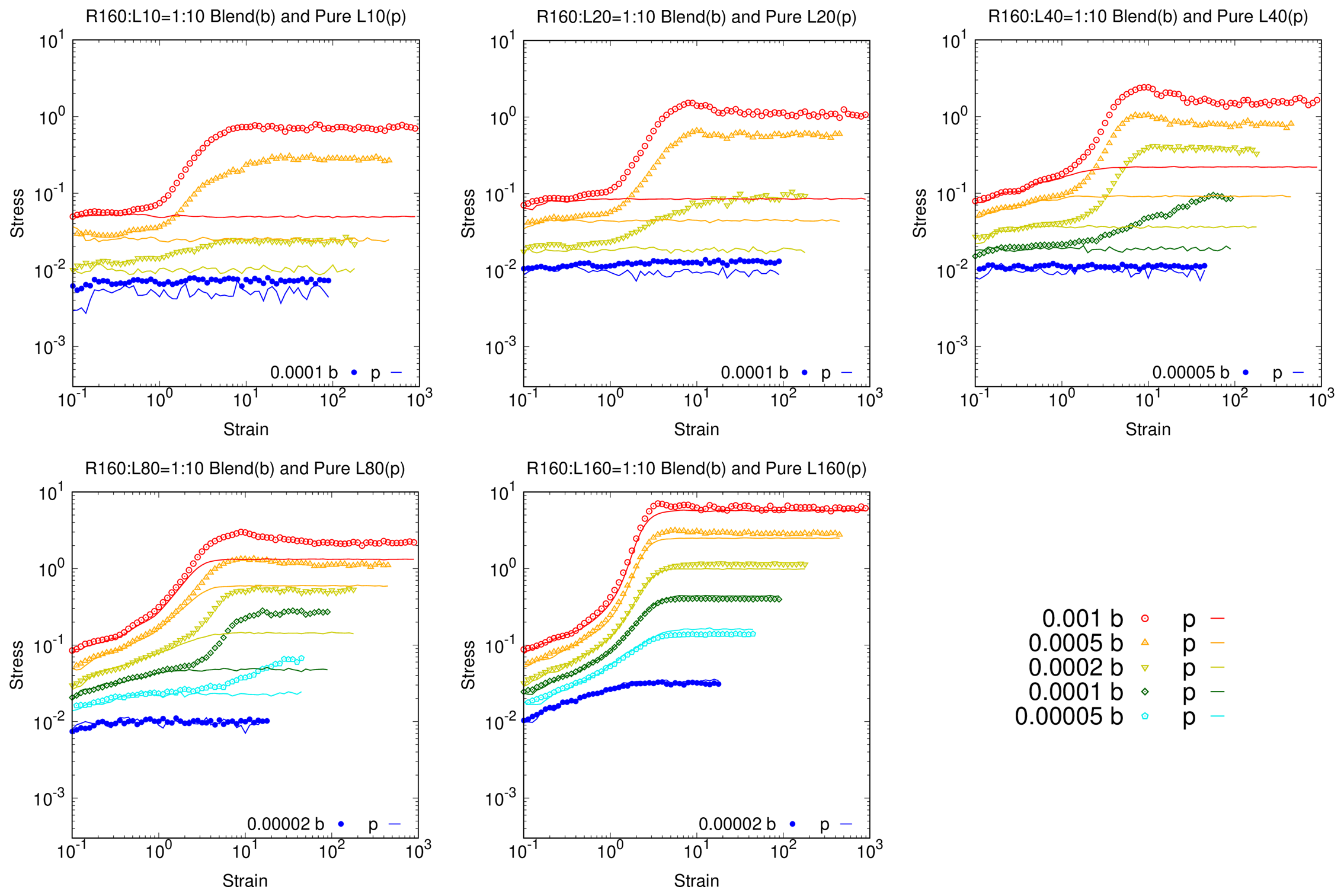}
    \caption{Stress--strain curves under biaxial elongational flows with biaxial elongational rates $\dot{\varepsilon}$. Symbols and lines correspond to those in Fig.~\ref{fig3_visc}.}
    \label{fig4_ss}
\end{figure*}

The viscosity growth curves $\eta_{\rm B}(t)$ and the stress--strain curves $\sigma_{\rm B}(\varepsilon)$ under biaxial elongational flows with the strain rate $\dot{\varepsilon}$ are represented in Figs.~\ref{fig3_visc} and \ref{fig4_ss}, respectively. 
The Savitzky--Golay filter~\cite{PressTeukolskyVetterlingFlannery2007} was used for smoothing the time series data of the biaxial elongational stress $\sigma_{\rm B}=(\sigma_{xx}+\sigma_{yy})/2-\sigma_{zz}$ and the biaxial elongational viscosity $\eta_{\rm B}=\sigma_{\rm B}/\dot{\varepsilon}$. 
Symbols and lines shown in Figs.~\ref{fig3_visc} and \ref{fig4_ss} represent the melts of ring-linear blends (b) and pure melts of linear chains (p), respectively.
The dashed lines represent the linear viscosity growth curve of ring-linear blends for biaxial elongational flow, $6\eta(t)$. 
The linear viscosity growth curve $\eta(t)$ is obtained from the time integration of the relaxation modulus $G(t)$:
\begin{align}
\eta(t)=\int_0^{t} G(t') {\rm d}t'. \label{eq.lin_visc}
\end{align}
The maximum strain rates overlapping with the linear viscosity growth curve $6\eta(t)$ are represented by filled circles in Figs.~\ref{fig3_visc} and \ref{fig4_ss}. We call these strain rates the threshold strain rates $\dot{\varepsilon}_{\rm th}$.
The data below the threshold strain rates show large fluctuations. 
To improve the visibility of the graph, they are not presented in Figs.~\ref{fig3_visc} and \ref{fig4_ss}.

When the strain rates are less than or equal to the threshold strain rates $\dot{\varepsilon}_{\rm th}$,
the viscosity and stress monotonically increase up to $\varepsilon=1$.
The steady states then appear at $\varepsilon > 1$.
When the strain rates are higher than the threshold strain rates $\dot{\varepsilon}_{\rm th}$, nonlinear behaviors are observed in ring-linear blends.
The strain hardening phenomenon, which is the nonlinear increase at a strain of approximately unity, are observed in the ring-linear blends and the weakly entangled pure linear melts (L80 and L160). In these melts, the steady states appear at  $\varepsilon > 10$ for R160-L10, R160-L20, R160-L40, and R160-L80, while at $\varepsilon < 10$ for L80 and L160, and R160-L160, as shown in Fig.~\ref{fig4_ss}.
The nonlinear behaviors of the pure linear melts, L80 and L160, are consistent with our previous work~\cite{MurashimaHagitaKawakatsu2018}. It is noted that the strain of 10 is very large and is not achievable in the current experimental techniques where the maximum strain in biaxial elongation is approximately 2~\cite{Venerus2019}.

In line with our expectations from the relaxation modulus shown in Fig.~\ref{fig2_gt}, R160-L10 and R160-L20 exhibit the strain hardening phenomenon owing to rings, because the two-step relaxation in Fig.~\ref{fig2_gt} arises from the ring components of the blends.
The steady-state viscosities in R160-L160 are weakly enhanced unlike the pure melt of L160, reflecting the small deviation between R160-L160 and L160 in Fig.~\ref{fig2_gt}. 

Contrary to our expectations from the relaxation modulus, R160-L40 and R160-L80 exhibit an apparent enhancement in steady-state viscosities compared to the pure linear melts L40 and L80, as shown in Fig.~\ref{fig3_visc}.  
Moreover, viscosity overshoot phenomena are observed when $\dot{\varepsilon}\ge 0.0005(\approx 1/\tau_{\rm e})$ in R160-L20, R160-L40, and R160-L80.
The viscosity overshoot in R160-L40 is more pronounced at higher strain rates. 

The strain hardening phenomena observed in R160-L40 and R160-L80 are not so anomalous because the small amount of long rings has a longer relaxation time than that of the host linear chains, and is expected to dominate the nonlinear behavior in the same way as a small amount of long linear chain~\cite{SugimotoMasubuchiTakimotoKoyama2001}.
We were surprised by the viscosity overshoot observed in R160-L20, R160-L40, and R160-L80, but not observed in R160-L10 and R160-L160. 
As R160-L10 does not show such overshoot phenomena, entanglements or “hooking” among ring and linear chains are important for the overshoot to occur. 
In R160-L160, however, the overshoot is suppressed (or is negligibly small). 
Thus, the overshoot phenomena found here are not observed in highly entangled melts.
In the more entangled melts, segments with approximately entanglement length $N_{\rm e} (\approx 70)$~\cite{Sukumaran2005} in long linear chains dominate the nonlinear behaviors in fast elongational flow.
The contribution from ring-linear interactions becomes relatively small in highly entangled melts.

Note that the threshold strain rates depend on the linear chain length $N_{\rm L}$, roughly estimated as $\dot{\varepsilon}_{\rm th}(N_{\rm L}) \sim N_{\rm L}^{-1/2}$, because
$\dot{\varepsilon}_{\rm th}=0.0001$ for R160-L10, $\dot{\varepsilon}_{\rm th}=0.00005$ for R160-L40, and $\dot{\varepsilon}_{\rm th}=0.00002$ for R160-L160.
The exponent $-1/2$ differs from the
conventional exponent $-2$ expected from the inverse of the Rouse relaxation time.
This curious relationship implies that the dynamics of ring polymers depend on the dynamics of linear chains.
The nonlinear viscosity growth and the strain hardening are not found in the pure linear melts of L10, L20, and L40, whereas they are found in L80 and L160. This suggests that the linear chains of L10, L20, and L40 do not contribute to the viscosity growth and strain hardening phenomena.  
If the small amount of long rings causes only the nonlinear strain hardening, the threshold strain rates $\dot{\varepsilon}_{\rm th}$ in R160-L10, R160-L20, and R160-L40 should be exactly the same, because the longest relaxation times in R160-L10, R160-L20, and R160-L40 are the same, corresponding to that in R160, as observed in Fig.~\ref{fig2_gt}. 
The threshold strain rate $\dot{\varepsilon}_{\rm th}$ should be determined by the relaxation time of R160.  
However, this conventional hypothesis does not hold in the ring-linear blends, as found here. Thus, the dependence of the threshold strain rate on the linear chain length arises from the cooperative phenomena of the ring and linear chains.

In Appendix \ref{App:mode}, we investigated the normal mode relaxation of ring polymers in ring-linear blends. It is surprising that the relaxation of ring polymers slows down as the linear chain length increases. The relaxation time of ring polymers $\tau_{2, \rm{R160}}$ depends on the linear chain length $N_{\rm L}$ as $\tau_{2,\rm{R160}}(N_{\rm L}) \sim N_{\rm L}^{1/2}$. Therefore, the curious dependence of the threshold strain rate $\dot{\varepsilon}(N_{\rm L}) \sim N_{\rm L}^{-1/2}$ was raised from the slow mode of ring polymers in the matrix of linear chains.

In the following sections, we focus on the viscosity overshoot phenomena, and would like to clarify its characteristics and the microscopic mechanism of viscosity overshoot in biaxial elongational flow.

\clearpage

\subsection{Average bond length of rings and linear chains in ring-linear blends under biaxial elongational flow}

Viscosity overshoot in biaxial elongational flows was observed when the strain rate is higher than $1/\tau_{\rm e}$, where $\tau_{\rm e} (\approx 2000)$~\cite{KremerGrest1990} is the entanglement time.
The entangled linear chains (L80 and L160) will be stretched when such a high strain rate is applied.
In the case of ring-linear blends, however, it is not obvious that the linear chains contributes to ring stretching.

\begin{figure*}[htbp]
    \centering
    \includegraphics[width=7in]{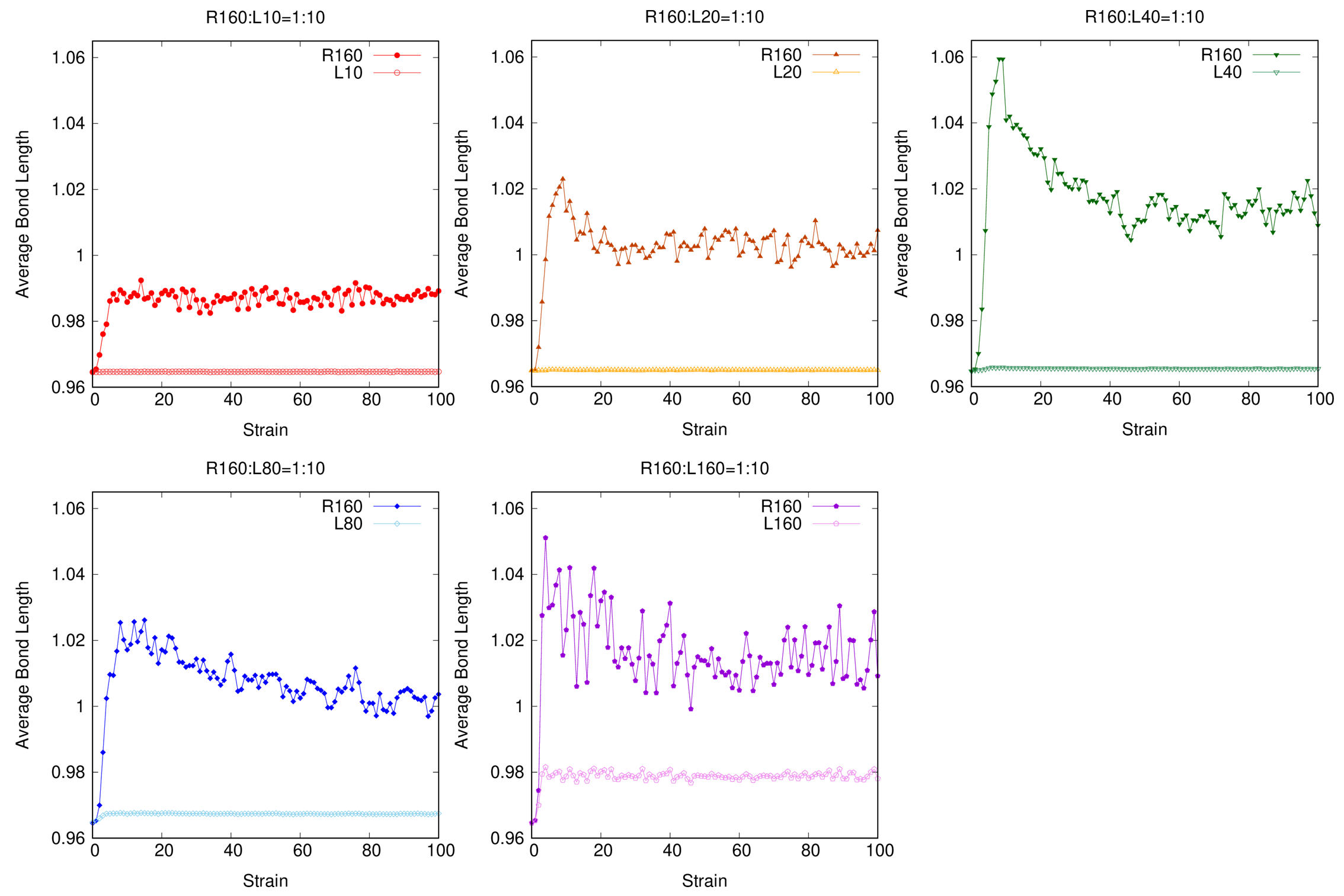}
    \caption{Average bond length of rings and linear chains in ring-linear blends under biaxial elongational flow with $\dot{\varepsilon}=0.001$. Filled and open symbols represent rings and linear chains, respectively. }
    \label{fig5_bond}
\end{figure*}

Figure \ref{fig5_bond} shows the average bond length of rings and linear chains in ring-linear blends under biaxial elongational flow with the strain rate $\dot{\varepsilon}=0.001$.
The equilibrium bond length is 0.965 for both rings and linear chains.
When applying strain, the bonds of rings start to stretch, whereas those of linear chains remain almost constant except for L80 and L160.
The overshoots of average bond length for rings are observed in ring-linear blends, except for R160-L10. 
Therefore, the viscosity overshoot could originate from the bond stretching of rings.

The bond stretching of rings is pronounced in R160-L40.
It is likely that the linear chains help to spread rings parallel to the elongational plane under biaxial elongational flow.
The longer chain contributes more to ring expansion, because the longer chain can coil around a ring and extend it by force from inside to outside in biaxial elongational flow. 
In the cases of R160-L80 and R160-L160 blends, however, the entangled network structure composed of linear chains prevents ring spreading.
The rings in R160-L80 and R160-L160 are prevented from expanding by the force from the linear chains outside of the rings.

Note that the temperature slightly increases in fast elongational flow at $\dot{\varepsilon} \ge 1/\tau_{\rm e}$, as summarized in Table \ref{table_temp}.
The bond stretching causes the temperature rise.
The rise is clearer in longer chain matrix especially for R160-L160.
Because the damping force in Eq.~\eqref{eq:force} is weak~\cite{KremerGrest1990}, the temperature is not sufficiently controlled under fast elongational flow.
The increase of the temperature accelerates the relaxation.
The relaxation of the overshoot becomes faster than that in the target temperature $T=1.0$.
So far, we have not quantitatively evaluated the influence of the temperature.
It can be negligibly small, because the temperature rise is up to 1.2 \% in the present scope.

\begin{table*}[htbp]
\centering
\caption{Temperature of ring-linear blends under biaxial elongational flow at $\dot{\varepsilon} \ge 1/\tau_{\rm e}$ in steady state.}
\begin{tabular}{c|rrrrr}
\hline
$\dot{\varepsilon}$    & R160-L10 & R160-L20 & R160-L40 & R160-L80 & R160-L160 \\
\hline
0.0005  & $1.0007\pm 0.0011$ & $1.0013\pm 0.0011$ & $1.0015\pm 0.0010$ & $1.0016\pm 0.0011$ & $1.0033\pm 0.0010$ \\
0.001   & $1.0028\pm 0.0013$ & $1.0035\pm 0.0013$ & $1.0043\pm 0.0012$ & $1.0057\pm 0.0012$ & $1.0120\pm 0.0011$ \\
\hline
\end{tabular}
\label{table_temp}
\end{table*}

\clearpage

\subsection{Direct observation of rings under biaxial elongational flow (R160-L40)}

\begin{figure*}[htbp]
    \centering
    \includegraphics[width=7in]{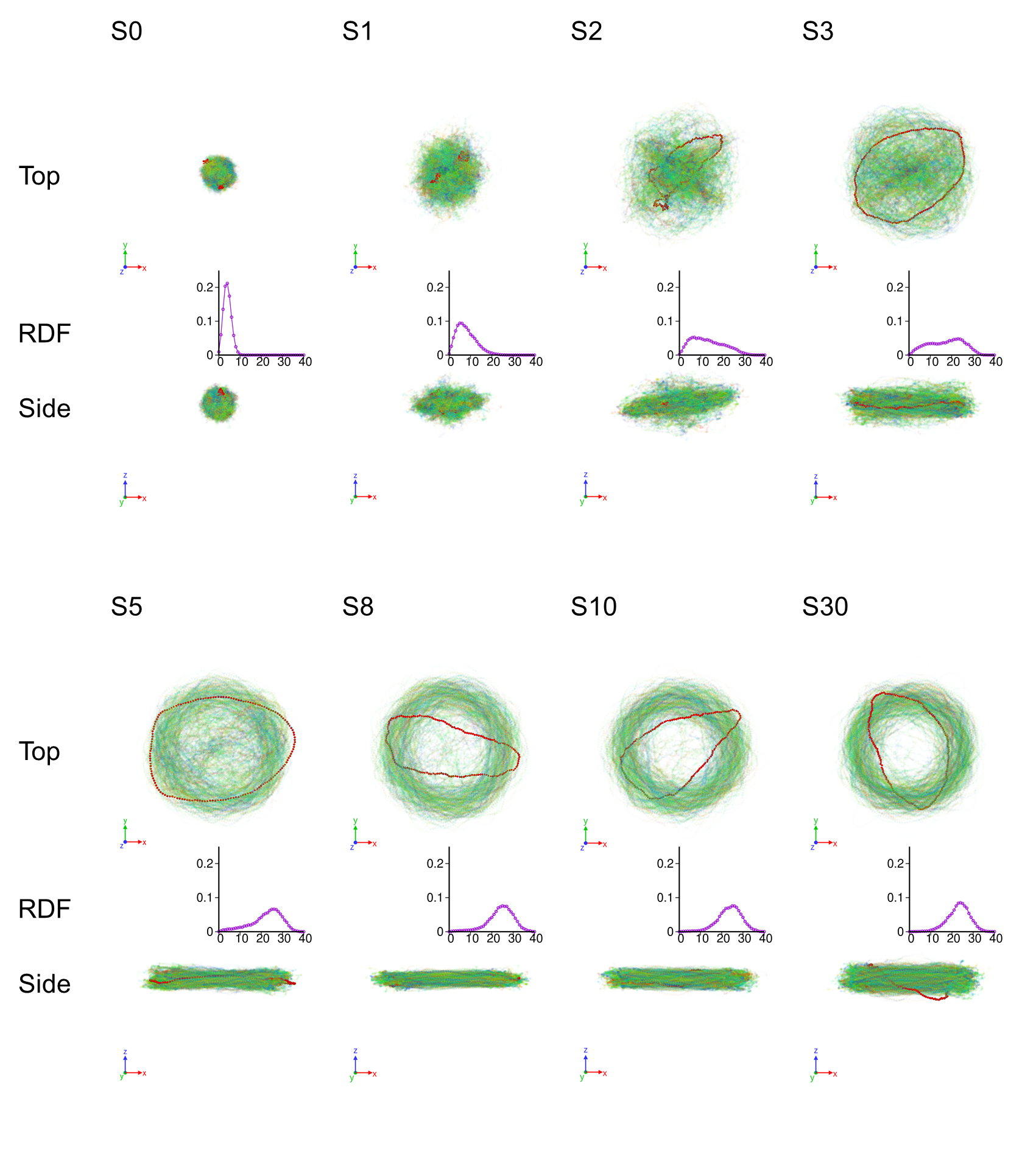}
    \caption{Snapshots of rings under biaxial elongational flow with $\dot{\varepsilon}=0.001$ in R160-L40. Top view (seeing $xy$-plane, or the elongational plane) and side view ($xz$-plane) are presented. Each column corresponds ${\rm S}n \, (n=0,1,2,3,5,8,10,30)$, representing $\varepsilon=n$. Thick rings with red color are identical. The graphs of the radial distribution function (RDF) are presented among the top and side views.}
    \label{fig6_ringshape}
\end{figure*}

Figure \ref{fig6_ringshape} shows snapshots of rings under biaxial elongational flow with $\dot{\varepsilon}=0.001$ in R160-L40. Each center of mass of a ring is set to the origin, and all rings ($M_{\rm R160}=384$) in R160-L40 are displayed. 
One of the rings is highlighted as a guide for the eye. 
These superposed rings represent the mean shape of the rings under biaxial elongational flow.

Focusing on the top and side views in Fig.~\ref{fig6_ringshape}, the size of the rings expands in the elongational plane with increasing strain $\varepsilon$, and the ring center opens at approximately $\varepsilon=8$, where the viscosity is maximum. 
For $\varepsilon > 8$, the size of the distribution in the elongational plane is slightly reduced, and the thickness of the distribution along the $z$-direction increases as the strain $\varepsilon$ increases. 
Because the size of the distribution in the elongational plane and the thickness of the distribution along the z-direction contribute positively and negatively to $\sigma_{\rm B}$, respectively, we can conclude that the viscosity overshoot observed in R160-L40 is caused by the dynamical behavior of rings under biaxial elongational flows. 
In small strains up to the maximum viscosity, the rings are gradually stretched in the elongational plane. 
After achieving the maximum viscosity, the rings stretched in the elongational plane relaxed slightly. To identify the reason of the relaxation of rings, we next investigate the number of linear chains penetrating through a ring.

\clearpage

\subsection{Probability distribution of number of linear chains penetrating through a ring (R160-L40)}

Now, we consider the number of linear chains penetrating through a ring, $n_{\rm p}$.
Accounting for all rings, we can obtain its distribution function $P(n_{\rm p})$.
To count the number of linear chains penetrating through a ring, we checked all pairs consisting of a ring and a linear chain, and judged whether the selected linear chain penetrates through the selected ring or not. This method has been proposed in our recent work~\cite{HagitaMurashima2021} and explained as follows. 
We create a loop from a linear chain by connecting both ends. 
This extra connection is always placed outside the ring. 
If the ring and the loop created from the linear chain form a catenane, the linear chain is judged to be penetrating through the ring.
We used Topoly~\cite{Topoly} for the judgement.

\begin{figure}[htbp]
    \centering
    \includegraphics[width=7in]{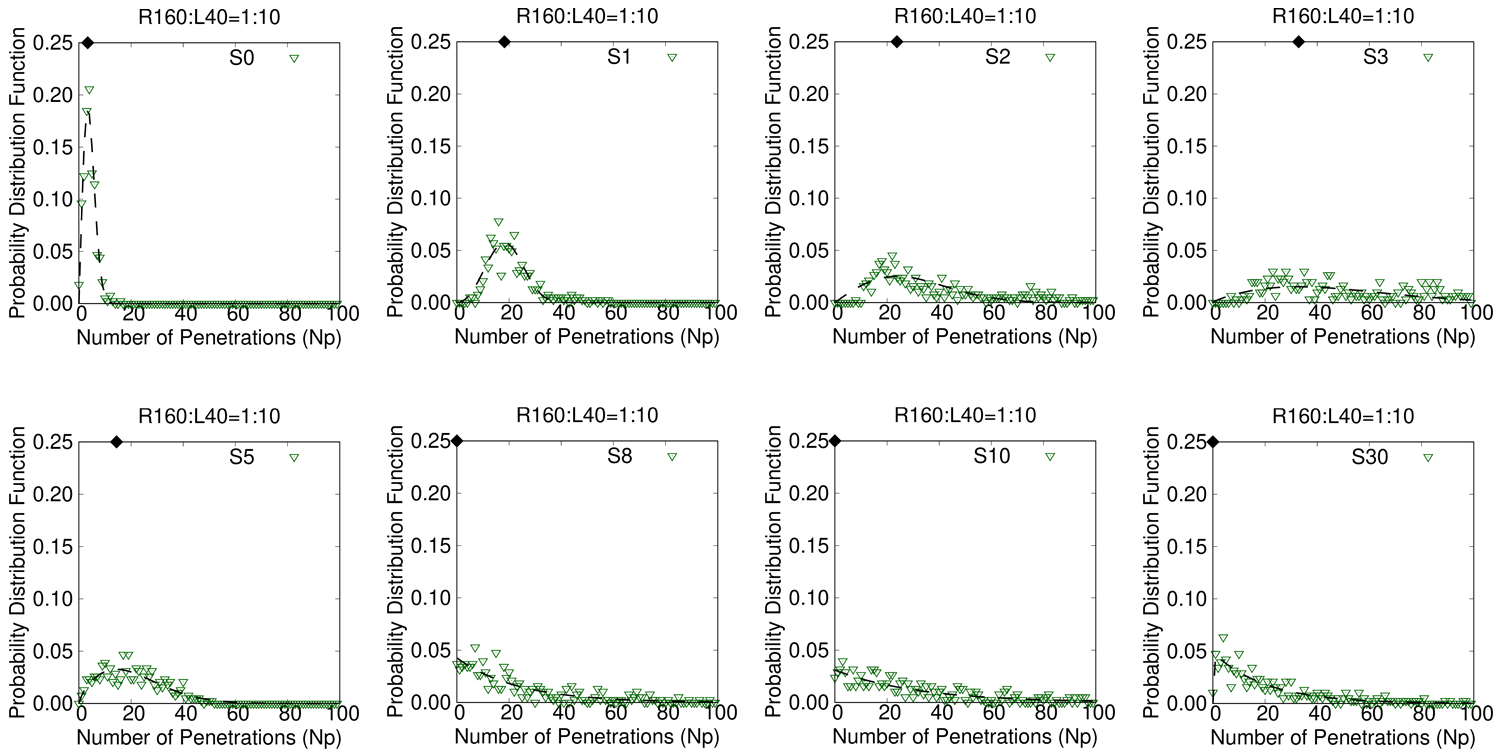}
    \caption{Probability distribution function of the number of penetrations $P(n_{\rm p})$ under biaxial elongational flow with $\dot{\varepsilon}=0.001$. ${\rm S}n \, (n=0,1,2,3,5,8,10,30)$ represents $\varepsilon=n$. Each dashed line, prepared as a guide for the eye, represents the Weibull distribution obtained from a nonlinear-least-squares curve fitting. The black diamond symbol shown at the top of each graph represents the peak position $n_{\rm p,peak}$.}
    \label{fig7_penetration}
\end{figure}

Figure \ref{fig7_penetration} shows the probability distribution function of the number of penetrations, $P(n_{\rm p})$, obtained from the configurations for $\varepsilon=\{0,1,2,3,5,8,10,30\}$ under biaxial elongational flow with $\dot{\varepsilon}=0.001$ in the case of R160-L40. 
We used the Weibull distribution function~\cite{Weibull1951}, $f(n_{\rm p},\lambda,\alpha)=\frac{\alpha}{\lambda}\left(\frac{n_{\rm p}}{\lambda} \right)^{\alpha-1} \exp \left(- \left(\frac{n_{\rm p}}{\lambda} \right)^{\alpha} \right)$, 
where $\alpha$ is the shape parameter and $\lambda$ is the scale parameter,
as a fitting function of $P(n_{\rm p})$ as a guide for the eye.
The Weibull distribution function interpolates the Rayleigh distribution ($\alpha>1$) and the exponential distribution ($\alpha=1$). The obtained parameters of the Weibull distribution function and the peak position $n_{\rm p,peak}=\lambda\left(1-\frac{1}{\alpha} \right)^{\frac{1}{\alpha}}$ of the Weibull distribution function are summarized in Table~\ref{table_weibull}.

As the strain is increased up to $\varepsilon=3$, the peak position shifts to the larger $n_{\rm p}$ side. 
At the same time, the height of the peak decreases as the variance of distribution. 
For $3 < \varepsilon < 8$, the peak position decreases to $n_{\rm p}=0$ as shown in Fig.~\ref{fig7_penetration} and Table~\ref{table_weibull}. 
When $\varepsilon \ge 8$, the probability distribution of $n_{\rm p}$ is suppressed in the larger $n_{\rm p}$ side and increased in the lower $n_{\rm p}$ side as the strain increases.
The shape of the distribution changes from unimodal (the Rayleigh distribution) to downslope (the exponential distribution) at $\varepsilon \approx 8$, where the viscosity is maximum. 
From these observations, it is certain that the linear chains inside the rings exit the rings when $\varepsilon > 8$.
In this situation, the stretched rings may shrink, 
because they lose the force originating from the linear chains that expanded the rings from inside to outside.

In Appendix \ref{App:penetration}, the distribution functions for R160-L10, R160-L20, R160-L80, and R160-L160 are presented.
Corresponding to the viscosity overshoot phenomena found in R160-L20 and R160-L80, in both cases the peak position of the Weibull distribution function goes to $n_{\rm p}=0$ at $\varepsilon=8$ as for R160-L40.
In the cases of R160-L10 and R160-L160, however, the peak position does not decrease to $n_{\rm p}=0$ while applying the strain.

\begin{figure}[htbp]
    \centering
    \includegraphics[width=3.5in]{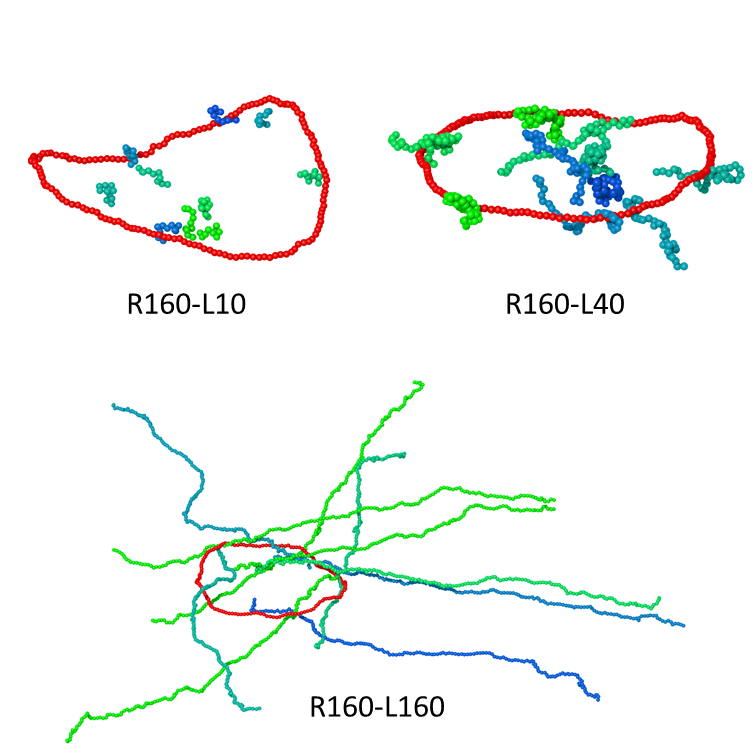}
    \caption{Snapshots at $\varepsilon=8$ in R160-L10, R160-L40, and R160-L160 with $n_{\rm p}=10$.}
    \label{fig8_ring1_line10}
\end{figure}

To deepen our understanding, typical snapshots of a ring and linear chains penetrating through the ring for R160-L10, R160-L40, and R160-L160 with $n_{\rm p}=10$ are shown in Fig.~\ref{fig8_ring1_line10}.
The rings open at $\varepsilon=8$.
The radius of the ring is much larger than the linear chains of L10 and L40.
It is easy to imagine the linear chains passing through a ring for the short-chain cases.
In the case of R160-L160, however, the linear chains are also expanded at $\varepsilon=8$, and the end-to-end distance of the linear chain is much larger than the radius of the ring.
Therefore, it is difficult for the linear chains penetrating into a ring to be released in the case of R160-L160.
It is noted that some linear chains are coiled around a ring in R160-L40.
These linear chains are thought to be responsible for bond stretching of the ring.

\begin{table*}[htbp]
\centering
\caption{Parameters of the Weibull distribution function ($\alpha, \lambda$) and the peak position $n_{\rm p,peak}$.}
\begin{tabular}{crrrrrrrr}
\hline
    & S0 & S1 & S2 & S3 & S5 & S8 & S10 & S30 \\
\hline
$\alpha$        & 2.07 & 3.04 & 1.98 & 1.77 & 1.71 & 1.0 & 1.0 & 1.0 \\
$\lambda$       & 4.66 & 20.78 & 33.9 & 52.7 & 24.1 & 23.5 & 32.3 & 20.5 \\
$n_{\rm p,peak}$& 3.39 & 18.2 & 23.8 & 32.9 & 14.4 & 0.0 & 0.0 & 0.0 \\
\hline
\end{tabular}
\label{table_weibull}
\end{table*}

\clearpage

\subsection{Comparisons with the other flow types (R160-L40)}

\begin{figure*}[htbp]
    \centering
    \includegraphics[width=7in]{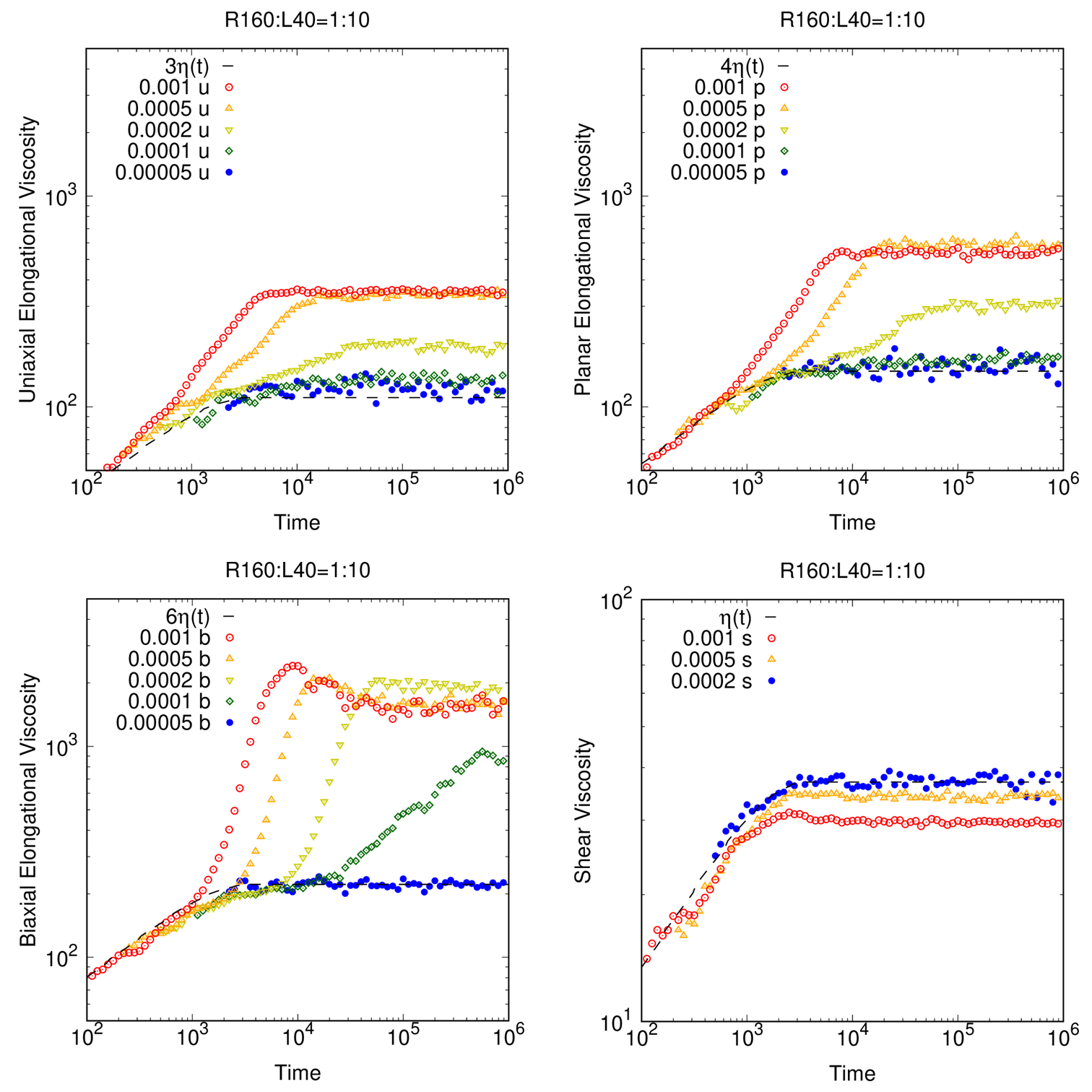}
    \caption{Viscosity growth curves under three typical (uniaxial, planar, biaxial) elongational flows, and shear flow in the case of R160-L40. The black dashed lines represent the linear viscosity growth curves; $3\eta(t)$, $4\eta(t)$, and $6\eta(t)$ for the uniaxial, planar, and biaxial elongational flows, respectively, and $\eta(t)$ for the shear flow.}
    \label{fig9_viscosity}
\end{figure*}

Finally, we summarized the viscosity growth curves under uniaxial, planar, biaxial elongational flows and shear flows in the case of R160-L40 in Fig.~\ref{fig9_viscosity}. The definitions of the viscosities, the stresses, and the strain rate tensors for uniaxial, planar elongational flows and shear flows are as follows: $\eta_{\rm U}=\sigma_{\rm U}/\dot{\varepsilon}$, $\eta_{\rm P}=\sigma_{\rm P}/\dot{\varepsilon}$, $\eta_{\rm S}=\sigma_{\rm S}/\dot{\gamma}$, $\sigma_{\rm U}=\sigma_{zz}-(\sigma_{xx}+\sigma_{yy})/2$, $\sigma_{\rm P}=\sigma_{zz}-\sigma_{xx}$, $\sigma_{\rm S}=\sigma_{xy}$, $\mathbf{K}={\rm diag}(-\dot{\varepsilon}/2,-\dot{\varepsilon}/2,\dot{\varepsilon})$ for uniaxial elongational flow, $\mathbf{K}={\rm diag}(-\dot{\varepsilon},0,\dot{\varepsilon})$ for planar elongational flow, and $\mathrm{K}_{xy}=\dot{\gamma}$ (the other components are zero) for shear flow. In the linear regimes, the viscosity growth curves correspond to the linear viscosity growth curves, $3\eta(t)$, $4\eta(t)$, and $6\eta(t)$ for uniaxial, planar, and biaxial elongational flows, respectively, and $\eta(t)$ for shear flows~\cite{Macosko,MurashimaHagitaKawakatsu2018}. 

As found in uniaxial and planar elongational flows, and in shear flow, the nonlinear behaviors, strain hardening in elongational flow and shear thinning in shear flow, are observed because the rings are longer than the host linear chains. 
The nonlinearity is, however, much more pronounced in biaxial elongational flow. 
We can conclude that the cooperative behavior of the rings and linear chains is enhanced under biaxial elongational flow.

It is noted that viscosity overshoot is not observed in uniaxial elongational flow as shown in Fig.~\ref{fig9_viscosity}.
Appendix \ref{App:uni} shows the uniaxial elongational viscosity for R160-L160 where the linear chains are weakly entangled.
We could not observe the viscosity overshoot under uniaxial elongational flow even in the weakly entangled ring-linear blend.
These results are different from the anomalous thickening~\cite{OConnorGeRubinsteinGrest2020} 
and the threading-unthreading transition~\cite{Borgeretal2020}
where the viscosity overshoot was observed in uniaxial elongational flow.
In the present study, there is no ring--ring interaction, because the ring fraction is smaller than the closest packing structure of rings.
Thus, the permanent threading~\cite{OConnorGeRubinsteinGrest2020} is not formed.
The threading-unthreading transition in uniaxial elongational flow was observed, when the ring fraction was higher than the overlap concentration of rings, and the rings were embedded in the well-entangled network~\cite{Borgeretal2020}.
It is likely that the interactions among the rings and the well-entangled network are important for the threading-unthreading transition in uniaxial elongational flow.

\clearpage

\section{Summary}
We investigated ring-linear blends under biaxial elongational flows by employing coarse-grained molecular dynamics simulation. 
We found cooperative phenomena of ring and linear chains in the nonlinear flow regimes. 
Viscosity overshoot phenomena were observed in the blends of R160-L20, R160-L40, and R160-L80, and were pronounced at the higher strain rates. 
Moreover, we found the curious relationship that the chain-length dependence of the threshold strain rate distinguishing the linear and nonlinear behaviors was estimated as $\dot{\varepsilon}_{\rm th}(N_{\rm L}) \sim N_{\rm L}^{-1/2}$. 
From the normal mode analysis, the relaxation time of ring polymers has the following chain-length dependence: $\tau_{2,\rm{R160}}(N_{\rm L}) \sim N_{\rm L}^{1/2}$.
Thus, the threshold strain rate is related to the relaxation time of ring polymers as $\dot{\varepsilon}_{\rm th}(N_{\rm L}) \sim 1/\tau_{2,\rm{R160}}(N_{\rm L})$.
To discover the cause of the stress overshoot, we directly observed the rings under biaxial elongational flows and investigated the number of linear chains penetrating through the rings. We found that the bond length and the size of the rings in the elongational plane increased at small strains up to the maximum viscosity, and then the rings stretched in the elongational plane were slightly relaxed, corresponding to the viscosity overshoot. 
The relaxation of the stretched rings is related to the number of penetrations. 
We concluded that the viscosity overshoot found in R160-L20, R160-L40, and R160-L80 under biaxial elongational flow is because the maximum stretching of ring polymers in the elongational plane and the relaxation of stretched rings caused by the release of linear chains penetrating through the rings.
The overshoot under biaxial elongational flow originates from the cooperative interactions between rings and linear chains.

From the comparisons among the different flow types (uniaxial, planar, biaxial elongational flows, and shear flow), the ring-linear blends show strong nonlinear behavior in biaxial elongational flow.
The ring polymers can be useful for developing a resilient material exposed to biaxial elongation or uniaxial compression.

The chain-length dependence of the threshold strain rate, $\dot{\varepsilon}_{\rm th}(N_{\rm L}) \sim N_{\rm L}^{-1/2}$, can be clarified by the usual experimental technique of biaxial elongation because it will be observed when the strain is approximately 1. No technical difficulties are associated with such small strains. 
The stress overshoot phenomena, however, are difficult to observe using the current experimental techniques, which cannot achieve strains over tens under biaxial elongation. Developments of new experimental techniques are highly demanded. Recent technological progress, such as the development of new rheometers~\cite{HuangMangnus2016}, has been remarkable. 
Our numerical predictions regarding the cooperative interactions between rings and linear chains are expected to be clarified in the future.
We hope our findings will boost the development of new material processing.

\begin{acknowledgement}
TM would like to thank Prof. T. Taniguchi, Prof. M. Sugimoto, Prof. J.-I. Takimoto, Prof. S. K. Sukumaran, and Prof. T. Uneyama, for fruitful discussions, comments, and encouragement. 
The authors thank Prof. H. Jinnai, Prof. T. Satoh, and Prof. T. Deguchi for their supports and encouragement. For the computations in this work, the authors were partially supported by the Supercomputer Center, the Institute for Solid State Physics, the University of Tokyo; MASAMUNE-IMR at the Center for Computational Materials Science, Institute for Materials Research, Tohoku University; Grand Chariot and Polaire at Hokkaido University Information Initiative Center; Flow at Nagoya University Information Technology Center; SQUID at Cybermedia Center, Osaka University; Fugaku at RIKEN Center for Computational Science; the Joint Usage/Research Center for Interdisciplinary Large-scale Information Infrastructures (JHPCN), and the High-Performance Computing Infrastructure (HPCI) in Japan: hp200048, hp200168, hp210132, jh210035. This work was partially supported by JSPS KAKENHI, Japan, Grant Numbers: JP18H04494, JP19H00905, JP20K03875, JP20H04649, and JST CREST, Japan, Grant Numbers: JPMJCR1993, JPMJCR19T4.
We would like to thank Editage (www.editage.com) for English language editing.
\end{acknowledgement}

\appendix 
\numberwithin{equation}{section}
\section{Generalized Kraynik--Reinelt boundary conditions under elongational flows\label{App:gKR}}

In a flow field $\mathbf{K}=(\boldsymbol{\nabla}\boldsymbol{v})^T$, the particles' positions and velocities are updated according to the SLLOD equation (Eqs.~\eqref{eq.SLLOD1} and \eqref{eq.SLLOD2}).
At the same time, the unit cell $\mathbf{L}=(\boldsymbol{e}_1,\boldsymbol{e}_2,\boldsymbol{e}_3)$, which is not necessarily equal to ${\rm diag}(L,L,L)$, is also deformed owing to the flow field $\mathbf{K}$:
\begin{equation}
\frac{d\mathbf{L}}{dt}=\mathbf{K}\cdot\mathbf{L}.
\end{equation}
Then, we find
\begin{equation}
\mathbf{L}(t)=\exp\left(\mathbf{K}t \right)\mathbf{L}(0)=(\boldsymbol{e}_1(t),\boldsymbol{e}_2(t),\boldsymbol{e}_3(t)).
\end{equation}
In general, the shape of $\mathbf{L}(t)$ is a parallelepiped. If a pair of parallel faces approach very close to each other and the gap between the faces is less than $r_{\rm c}$, the simulation will collapse. If we can find a unit cell $\tilde{\mathbf{L}}=(\tilde{\boldsymbol{e}}_1,\tilde{\boldsymbol{e}}_2,\tilde{\boldsymbol{e}}_3)$, that is close to a cubic shape or a rectangular parallelepiped, and the unit cell $\mathbf{L}(t)$ satisfies
\begin{equation}
\mathbf{L}(t)=(n_1 \tilde{\boldsymbol{e}}_1, n_2 \tilde{\boldsymbol{e}}_2, n_3 \tilde{\boldsymbol{e}}_3),\quad (n_i \in Z)
\end{equation}
then we can reduce $\mathbf{L}(t)$ to $\tilde{\mathbf{L}}$. The LLL--algorithm~\cite{LenstraLenstraLovasz1982} and Semaev’s algorithm~\cite{Semaev2001} are useful for finding $\tilde{\mathbf{L}}$.

The flow field $\mathbf{K}$ of biaxial elongational flow with elongational rate $\dot{\varepsilon}$ is represented as $\mathbf{K}={\rm diag}(\dot{\varepsilon}, \dot{\varepsilon}, -2\dot{\varepsilon})$.
Then, we find
\begin{equation}
\mathbf{L}(t)={\rm diag}\left( e^{\dot{\varepsilon}t}, e^{\dot{\varepsilon}t},  e^{-2\dot{\varepsilon}t} \right)\mathbf{L}(0).    
\end{equation}
If we choose $\mathbf{L}(0)={\rm diag}(L,L,L)$ as usual, the unit cell
$\mathbf{L}(t)={\rm diag}\left(L e^{\dot{\varepsilon}t}, L e^{\dot{\varepsilon}t}, L e^{-2\dot{\varepsilon}t} \right)$
 is flattened for finite time steps so that $\tilde{\mathbf{L}}$ cannot be found. 
 To avoid this, we choose an initial unit cell as~\cite{NicholsonRutledge2016}
 \begin{equation}
\mathbf{L}(0)=\mathbf{V}^{-1}=
\begin{pmatrix}
0.737 & 0.591 & 0.328 \\
-0.328 & 0.737 & -0.591 \\
-0.591 & 0.328 & 0.737
\end{pmatrix}.
 \end{equation}
The column vectors of $\mathbf{V}^{-1}$ are the eigenvectors of automorphism $\mathbf{M}\in {\rm SL}(3;Z)$ such that $\mathbf{L}(t+t_0)=\mathbf{M}\cdot\mathbf{L}(t)$. Here, ${\rm SL}(3;Z)$ denotes the special linear group over integers of degree three, which is the group of volume- and orientation-preserving linear transformations of $Z_3$. In this case, we can find a reduced unit cell $\tilde{\mathbf{L}}$, and then we continue to apply elongational flow to the system for very long duration, where the strain exceeds 10.

In practice, it is convenient to choose an upper triangular unit cell $\mathbf{L}_{\rm MD}$ in MD simulation utilizing spatial-decomposition techniques for parallel computing~\cite{HansenEvans1994}. We can convert $\mathbf{L}=(\boldsymbol{a},\boldsymbol{b},\boldsymbol{c})$ to $\mathbf{L}_{\rm MD}$ as
\begin{equation}
\mathbf{L}_{\rm MD}=
\begin{pmatrix}
\boldsymbol{a}\cdot\widehat{\boldsymbol{e}}_1 &
\boldsymbol{b}\cdot\widehat{\boldsymbol{e}}_1 &
\boldsymbol{c}\cdot\widehat{\boldsymbol{e}}_1 \\
0 & \boldsymbol{b}\cdot\widehat{\boldsymbol{e}}_2 & \boldsymbol{c}\cdot\widehat{\boldsymbol{e}}_2 \\
0 & 0 & \boldsymbol{c}\cdot\widehat{\boldsymbol{e}}_3
\end{pmatrix},
\end{equation}
where $\widehat{\boldsymbol{e}}_1 = \widehat{\boldsymbol{a}}=\boldsymbol{a}/|\boldsymbol{a}|$, $\widehat{\boldsymbol{e}}_2=(\widehat{\boldsymbol{a}}\times\widehat{\boldsymbol{b}})\times \widehat{\boldsymbol{a}}$,
and
$\widehat{\boldsymbol{e}}_3=\widehat{\boldsymbol{a}}\times\widehat{\boldsymbol{b}}$.
In other words, $\mathbf{L}$ and $\mathbf{L}_{\rm MD}$ are related through a rotation matrix $\mathbf{Q}$: 
$\mathbf{L}=\mathbf{Q}\cdot\mathbf{L}_{\rm MD}$~\cite{MurashimaUrataLi2019}. The rotation matrix $\mathbf{Q}$ is obtained by $\mathbf{Q}=\mathbf{L}\cdot\mathbf{L}_{\rm MD}^{-1}$. 
The flow field $\mathbf{K}$ is applied on the $\mathbf{L}$-frame, and then the positions, velocities and forces are updated on the $\mathbf{L}_{\rm MD}$-frame in a simulation algorithm~\cite{MurashimaHagitaKawakatsu2018}.

\clearpage

\section{Storage and loss moduli of pure linear chain melts, pure ring melts, and ring-linear blends\label{App:G1G2}}

\begin{figure*}[htbp]
    \centering
    \includegraphics[width=3.5in]{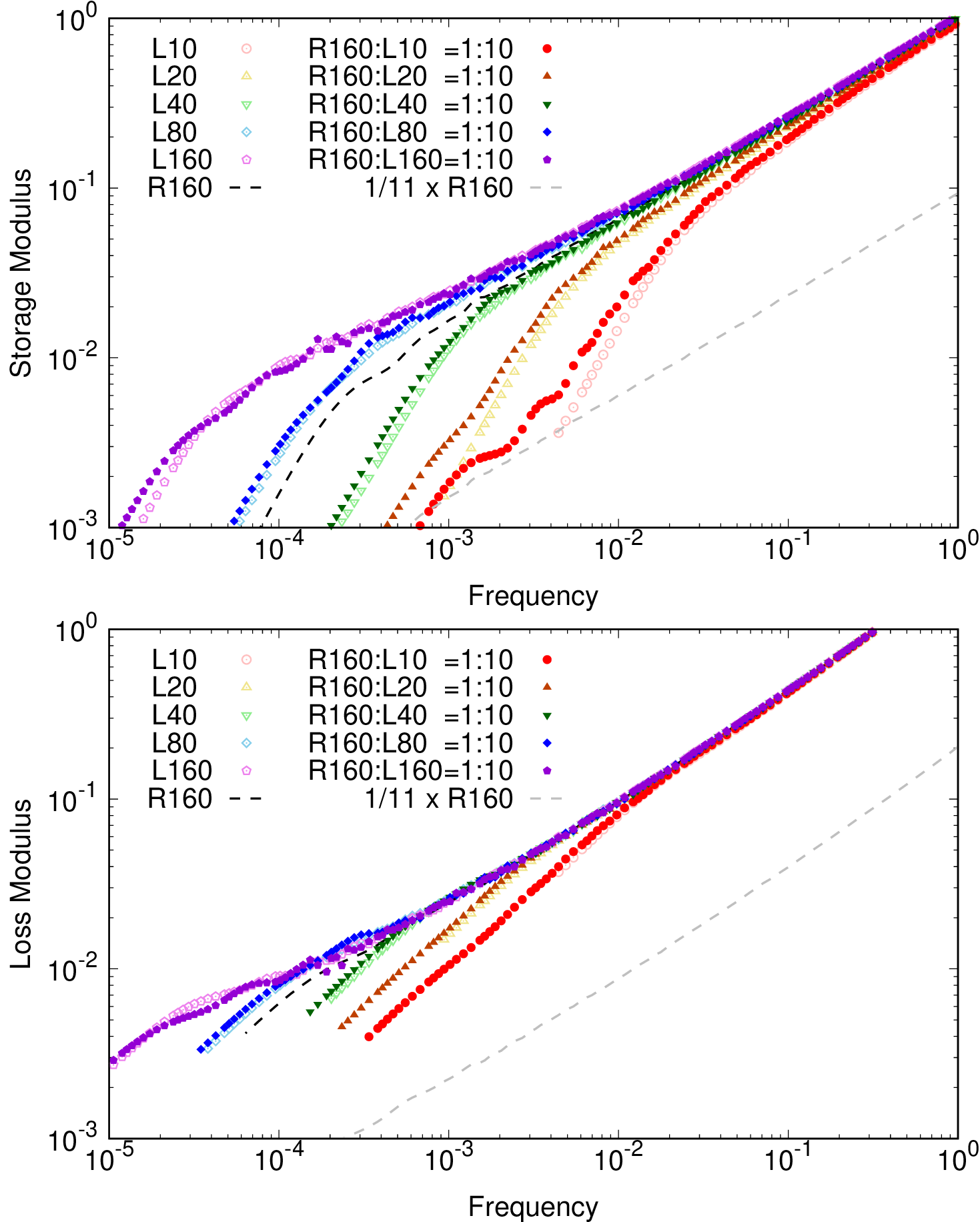}
    \caption{Comparison of storage (top) and loss (bottom) moduli of the pure linear chain melts (L10, L20, L40, L80, and L160), the pure ring melt (R160), and the ring-linear blends (R160-L10, R160-L20, R160-L40, R160-L80, and R160-L160).}
    \label{fig10_g1g2}
\end{figure*}

Figure \ref{fig10_g1g2} summarizes the storage and loss moduli of the pure linear chains (L10, L20, L40, L80, and L160), the pure ring melt (R160), and the ring-linear blends (R160-L10, R160-L20, R160-L40, R160-L80, and R160-L160). To obtain these moduli, we executed the following procedure.

In principle, the relaxation modulus $G(t)$ is converted to the storage and loss moduli, $G'(\omega)$ and $G''(\omega)$,  through the Fourier transform of $G(t)$, $\widehat{G}(\omega)$, as
\begin{equation}
    G'(\omega) + {\rm i} G''(\omega) = {\rm i} \omega \widehat{G}(\omega),
\end{equation}
where $\widehat{G}(\omega)=\int_{-\infty}^{\infty} G(t) e^{-{\rm i} \omega t} {\rm d}t$.
The relaxation modulus $G(t)$ was sampled over a finite time window and was the discrete data $G(t_k), k=\{1, \cdots, N\}$.
The Fourier transformation of $G(t)$ is problematic as it is.
To apply the Fourier transformation to the relaxation modulus $G(t)$ with a finite time window, the i-Rheo algorithm has been developed~\cite{Tassieri2016}:
\begin{align}
    -\omega^2 \widehat{G}(\omega) &= {\rm i} \omega G(0) 
    +\left( 1 - e^{-{\rm i} \omega t_1} \right)
    \frac{G(t_1)-G(0)}{t_1} \notag \\
    &+\dot{G}_{\infty} e^{-{\rm i}\omega t_N}
    + \sum_{k=2}^N 
    \left( 
    \frac{G(t_k)-G(t_{k-1})}{t_k - t_{k-1}}
    \right)
    \left(
    e^{-{\rm i}\omega t_{k-1}} - e^{-{\rm i}\omega t_k}
    \right),\label{eq:iRheo}
\end{align}
where $\displaystyle \dot{G}_{\infty}=\lim_{t\to\infty} \frac{{\rm d}G(t)}{{\rm d}t}$.
This algorithm is applicable for unequally spaced data.
Applying Eq.~\eqref{eq:iRheo} to $G(t)$, we obtained $G'(\omega)$ and $G''(\omega)$ as shown in Fig.~\ref{fig10_g1g2}.

\clearpage

\section{Normal mode analysis of ring polymers in ring-linear blends\label{App:mode}}

In this appendix, we discuss the normal mode of ring polymers in ring-linear blends.
The normal mode of a ring polymer composed of $N$ beads is represented with $p$ modes as~\cite{Tsolou2010}

\begin{align}
    \mathbf{Y}_p (t)= \frac{1}{N} \int_0^N \mathbf{R}_n (t) \sin\left(\frac{p\pi n}{N}\right) \rm{d}n,
\end{align}
where $\mathbf{R}_n$ represents the position of the $n$-th bead in a ring polymer ($0\le n < N$) and $p=\{2, 4, 6, \cdots\}$. 
The odd $p$-modes are suppressed in the normal mode of the ring polymer because of the ring periodicity, $\mathbf{R}_i=\mathbf{R}_{N+i}$.
Because the second mode ($p=2$) has the longest relaxation time in a ring polymer,
we calculate the autocorrelation function of the second mode $\langle \mathbf{Y}_2(t) \mathbf{Y}_2(0)\rangle$, and then estimate its relaxation time $\tau_2$.
The autocorrelation function of the second mode is approximated with a stretched exponential Kohlrausch--Williams--Watts (KWW) function as follows:
\begin{align}
    \frac{\langle \mathbf{Y}_2 (t) \cdot \mathbf{Y}_2(0) \rangle}{\langle \mathbf{Y}_2 (0)^2 \rangle} \approx \exp\left[ -\left( \frac{t}{\alpha_2} \right)^{\beta_2} \right].
\end{align}
The relaxation time of the second mode $\tau_2$ is determined from the characteristic relaxation time $\alpha_2$ and the stretching exponent $\beta_2$ as
\begin{align}
    \tau_2 = \frac{\alpha_2}{\beta_2} \Gamma\left(\frac{1}{\beta_2} \right). \label{eq.tau2}
\end{align}

\begin{figure*}[htbp]
    \centering
    \includegraphics[width=3.5in]{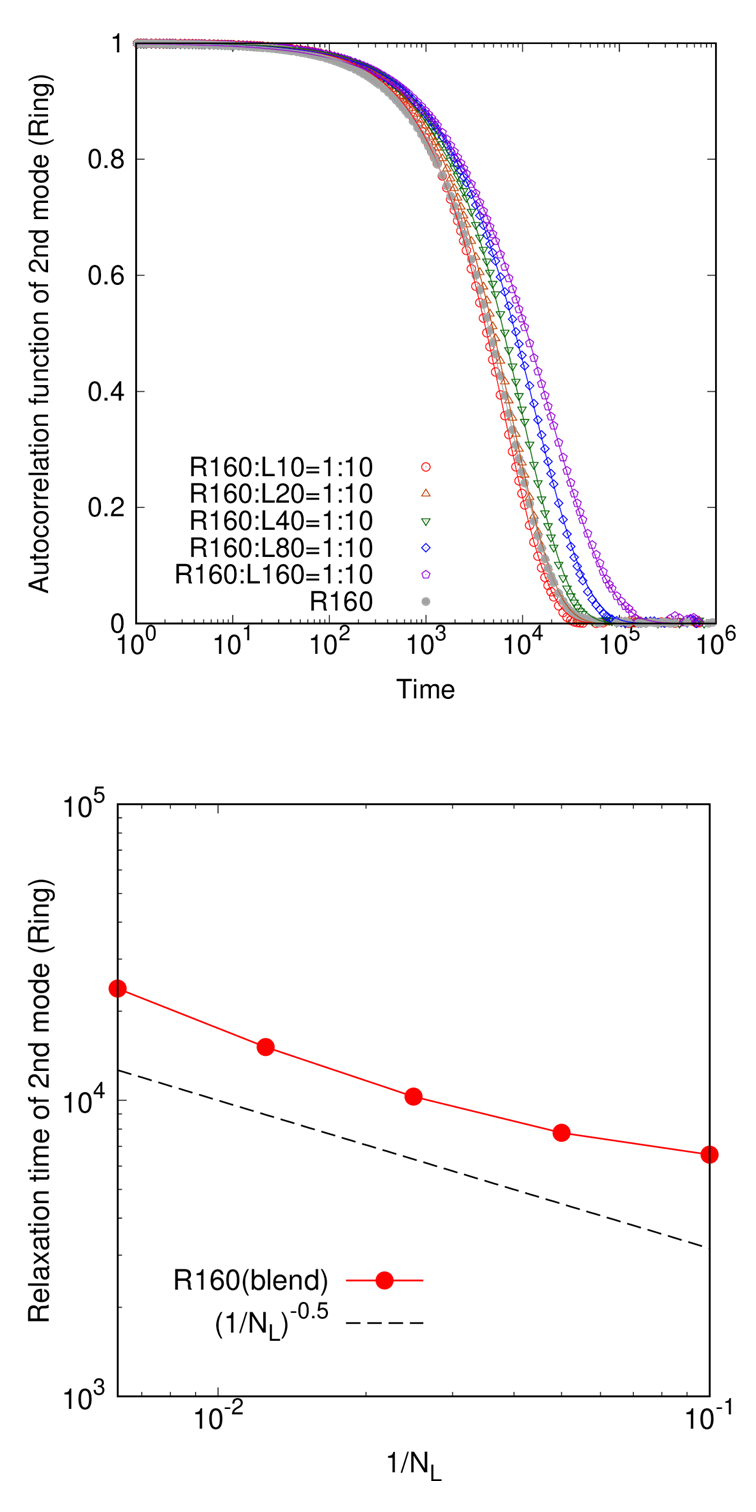}
    \caption{Autocorrelation function of the second mode for rings (top) and the chain-length dependency of the relaxation time $\tau_{2,\rm{R160}}$ (bottom).}
    \label{fig11_mode}
\end{figure*}

Figure \ref{fig11_mode} represents the autocorrelation functions of the second mode for ring  polymers in R160-L10, R160-L20, R160-L40, R160-L80, and R160-L160 blends and the obtained relaxation time for ring polymers with $N=160$, $\tau_{2,\rm{R160}}$.
For comparison, the autocorrelation function for the pure ring melt (R160) is also presented.
The parameters $\alpha_2$ and $\beta_2$ were determined from nonlinear-least-squares curve fitting.
Table~\ref{table_mode} summarizes the fitting parameters ($\alpha_2, \beta_2$) and the relaxation time $\tau_{2,\rm{R160}}$ obtained from Eq.~\eqref{eq.tau2}.
As shown in these results,
the relaxation time $\tau_{2,\rm{R160}}$ depends on the linear chain length, and the relationship found is $\tau_{2,\rm{R160}}(N_{\rm L})\sim N_{\rm L}^{1/2}$.

\begin{table*}[htbp]
\centering
\caption{Parameters of Kohlrausch--Williams--Watts (KWW) functions ($\alpha_2, \beta_2$) and relaxation time of ring polymers $\tau_{2,\rm{R160}}$ in ring-linear blends and pure ring melt.}
\begin{tabular}{crrrrrr}
\hline
    & R160-L10 & R160-L20 & R160-L40 & R160-L80 & R160-L160 & R160 \\
\hline
$\alpha_2$ & 6330 & 7500 & 9730 & 13300 & 18800 & 7020 \\
$\beta_2$ & 0.928 & 0.927 & 0.890 & 0.799 & 0.700 & 0.868 \\
$\tau_{2,\rm{R160}}$ & 6560 & 7770 & 10300 & 15100 & 23800 & 7540 \\
\hline
\end{tabular}
\label{table_mode}
\end{table*}

\clearpage

\section{Distribution function of number of linear chains penetrating through a ring for R160-L10, R160-L20, R160-L80, and R160-L160 \label{App:penetration}}

To facilitate the reader's understanding, this appendix presents graphs of the distribution function of the number of penetrations for R160-L10, R160-L20, R160-L80, and R160-L160.
The fitting parameters and the peak position of the Weibull distribution function are summarized in Table~\ref{table_weibull_all}.

\begin{table*}[htbp]
\centering
\caption{Parameters of the Weibull distribution function ($\alpha, \lambda$) and the peak position $n_{\rm p,peak}$.}
\begin{tabular}{crrrrrrrr}
\multicolumn{9}{c}{R160-L10} \\
\hline
    & S0 & S1 & S2 & S3 & S5 & S8 & S10 & S30 \\
\hline
$\alpha$        & 2.17 & 2.43 & 2.42 & 2.34 & 1.84 & 1.15 & 1.13 & 1.04 \\
$\lambda$       & 8.46 & 40.5 & 60.8 & 61.5 & 37.8 & 37.0 & 47.6 & 37.4 \\
$n_{\rm p,peak}$& 6.36 & 32.5 & 48.7 & 48.5 & 24.6 & 6.28 & 7.13 & 1.83 \\
\hline
\multicolumn{9}{c}{ } \\
\multicolumn{9}{c}{R160-L20} \\
\hline
    & S0 & S1 & S2 & S3 & S5 & S8 & S10 & S30 \\
\hline
$\alpha$        & 2.12 & 2.70 & 2.31 & 2.06 & 2.30 & 1.0 & 1.09 & 1.21 \\
$\lambda$       & 5.87 & 27.9 & 49.6 & 58.1 & 32.2 & 34.8 & 38.7 & 31.6 \\
$n_{\rm p,peak}$& 4.35 & 23.6 & 38.8 & 42.1 & 25.1 & 0.0 & 3.85 & 7.43 \\
\hline
\multicolumn{9}{c}{ } \\
\multicolumn{9}{c}{R160-L40} \\
\hline
    & S0 & S1 & S2 & S3 & S5 & S8 & S10 & S30 \\
\hline
$\alpha$        & 2.07 & 3.04 & 1.98 & 1.77 & 1.71 & 1.0 & 1.0 & 1.0 \\
$\lambda$       & 4.66 & 20.78 & 33.9 & 52.7 & 24.1 & 23.5 & 32.3 & 20.5 \\
$n_{\rm p,peak}$& 3.39 & 18.2 & 23.8 & 32.9 & 14.4 & 0.0 & 0.0 & 0.0 \\
\hline
\multicolumn{9}{c}{ } \\
\multicolumn{9}{c}{R160-L80} \\
\hline
    & S0 & S1 & S2 & S3 & S5 & S8 & S10 & S30 \\
\hline
$\alpha$        & 2.11 & 3.07 & 2.22 & 1.88 & 1.81 & 1.0 & 1.0 & 1.0 \\
$\lambda$       & 4.42 & 19.3 & 32.2 & 62.5 & 33.8 & 22.8 & 31.0 & 22.0 \\
$n_{\rm p,peak}$& 3.25 & 16.9 & 24.6 & 41.8 & 21.6 & 0.0 & 0.0 & 0.0 \\
\hline
\multicolumn{9}{c}{ } \\
\multicolumn{9}{c}{R160-L160} \\
\hline
    & S0 & S1 & S2 & S3 & S5 & S8 & S10 & S30 \\
\hline
$\alpha$        & 1.87 & 2.93 & 2.68 & 2.58 & 1.75 & 1.16 & 1.18 & 1.20 \\
$\lambda$       & 3.86 & 16.8 & 28.1 & 54.2 & 48.9 & 28.6 & 21.8 & 19.6 \\
$n_{\rm p,peak}$& 2.56 & 14.6 & 23.6 & 44.8 & 30.1 & 5.29 & 4.34 & 4.38 \\
\hline
\end{tabular}
\label{table_weibull_all}
\end{table*}

\begin{figure}[htbp]
    \centering
    \includegraphics[width=7in]{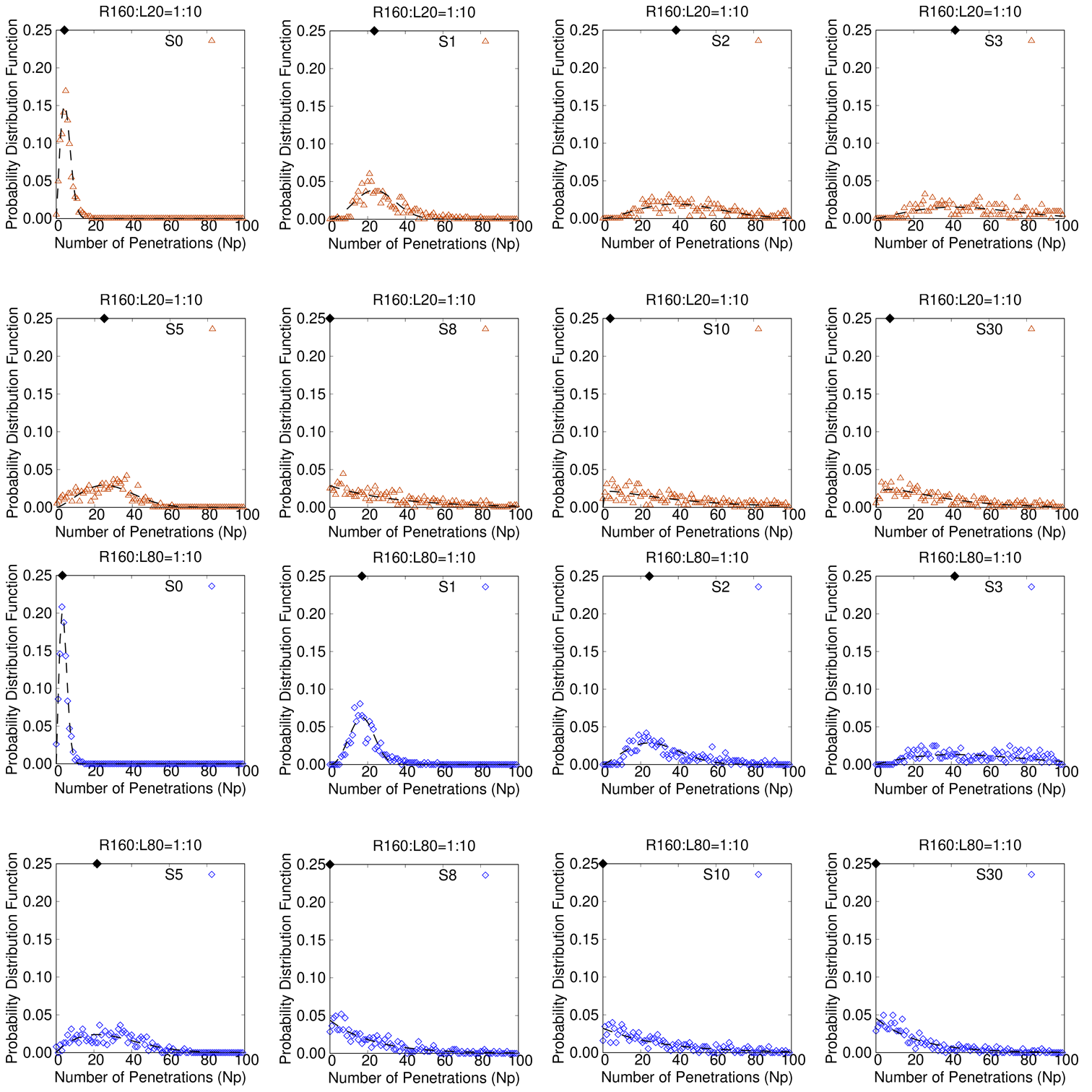}
    \caption{Probability distribution function of the number of penetrations $P(n_{\rm p})$ under biaxial elongational flow with $\dot{\varepsilon}=0.001$ for R160-L20 and R160-L80.}
    \label{fig12_penetration_20_80}
\end{figure}

In the cases of R160-L20 and R160-L80, the graphs show similarities with the R160-L40 graph.
These graphs are summarized in Fig.~\ref{fig12_penetration_20_80}.
In the case of R160-L20, however, the shapes of the distribution function are the Rayleigh distribution ($n_{\rm p,peak} > 0$) at $\varepsilon>8$.
These contributions are probably minor to the bond stretching, because the chain length $N_{\rm L}=20$ is short.

\begin{figure}[htbp]
    \centering
    \includegraphics[width=7in]{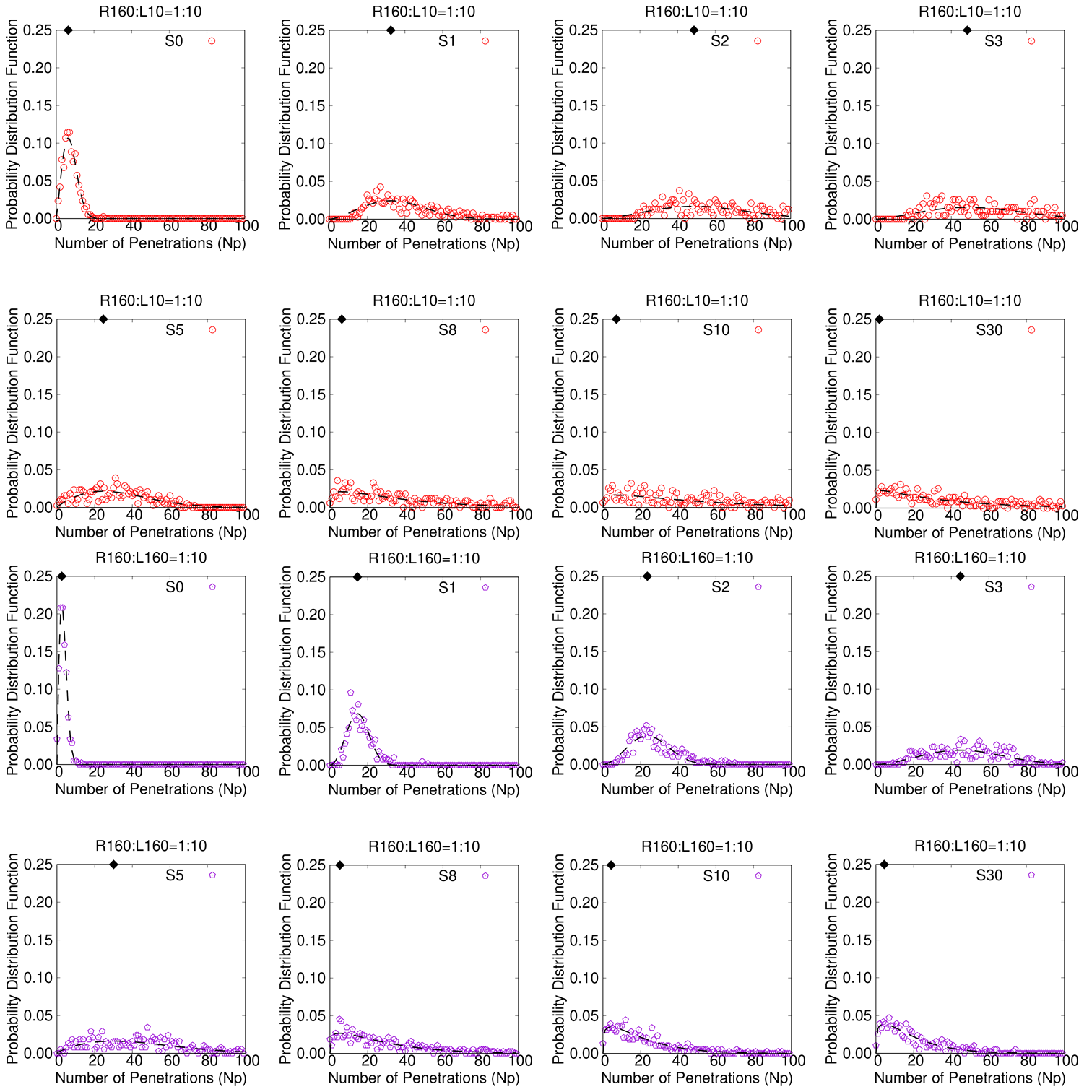}
    \caption{Probability distribution function of the number of penetrations $P(n_{\rm p})$ under biaxial elongational flow with $\dot{\varepsilon}=0.001$ for R160-L10 and R160-L160.}
    \label{fig13_penetration_10_160}
\end{figure}

The graphs for R160-L10 and R160-L160 are displayed in Fig.~\ref{fig13_penetration_10_160}.
In the case of R160-L10, the linear chain penetration does not contribute to the nonlinear behavior of the stress--strain and viscosity growth curves.
The shape of the distribution function exhibits a broad dispersion for R160-L10.
In the case of R160-L160 too, the graphs are similar to the R160-L40 graph up to $\varepsilon=5$.
For $\varepsilon \ge 8$, however, the peak position of the distribution function is not equal to zero.
This means that the release of linear chains penetrating through a ring is insufficient in R160-L160, because the end-to-end distance of linear chain is longer than the ring size as shown in Fig.~\ref{fig8_ring1_line10}.

\clearpage

\section{Uniaxial elongational viscosity for weakly entangled melt (R160-L160) \label{App:uni}}

\begin{figure}[htbp]
    \centering
    \includegraphics[width=3.5in]{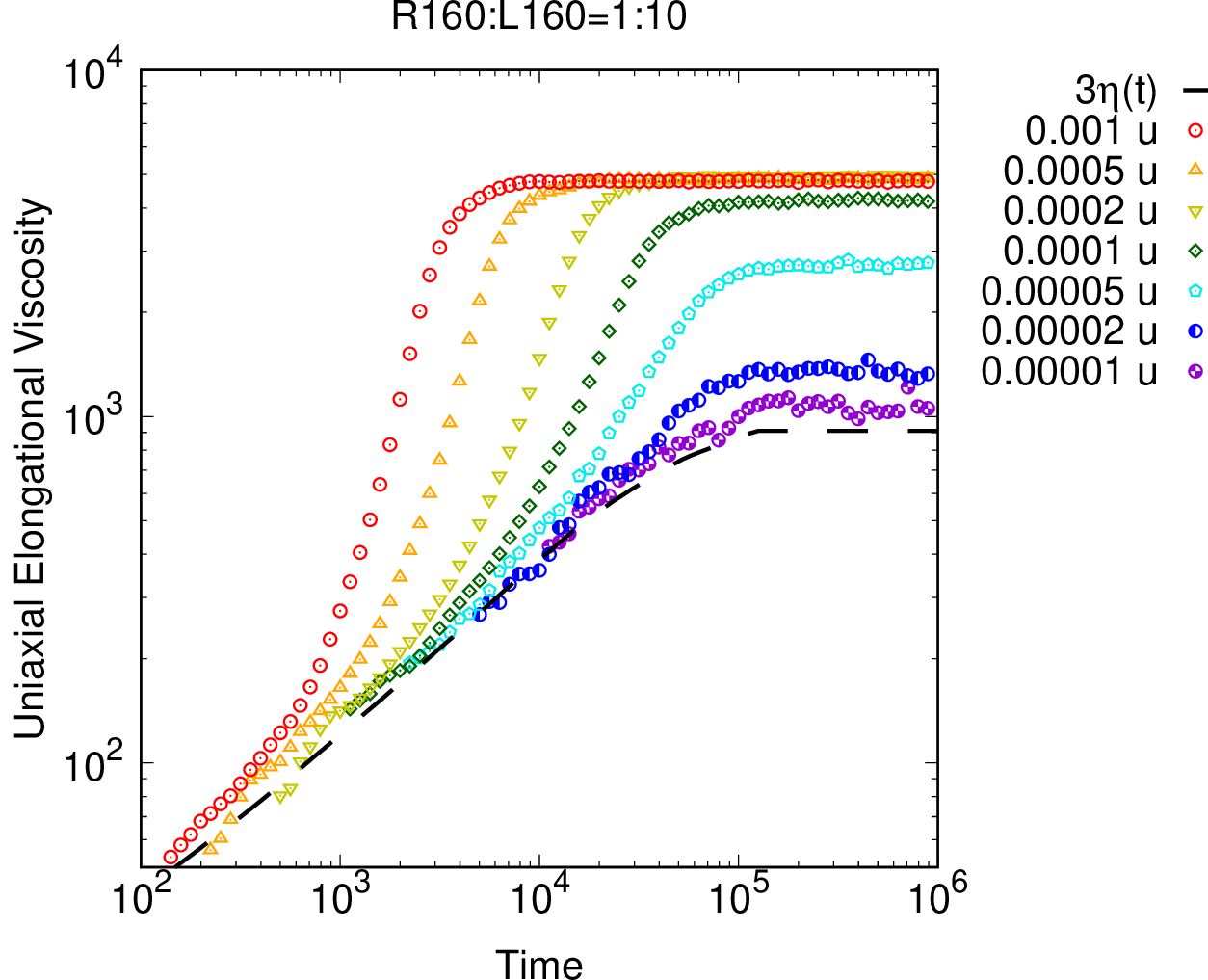}
    \caption{Uniaxial elongational viscosity for R160-L160.}
    \label{fig14_uniaxial}
\end{figure}

For comparison with the threading-unthreading transition under uniaxial elongational flow~\cite{Borgeretal2020}, the uniaxial elongational viscosity for R160-L160 where the rings are embedded in the weakly entangled linear chain network is presented in Fig.~\ref{fig14_uniaxial}.
The viscosity overshoot under uniaxial elongational flow is not observed in the present scope.
The threading-unthreading transition under uniaxial elongational flow was observed in R400:L400 = 3:7 using a bending potential to decrease $N_{\rm e}(\approx 28)$~\cite{Borgeretal2020}.
In the present work, we investigated flexible chains with $N_{\rm e}\approx 70$.
To observe the viscosity overshoot under uniaxial elongational flow in ring-linear blends, much entanglements and higher fraction of rings are needed.
The viscosity overshoot under biaxial elongational flow in the present study was observed in the short-chain matrices.
The mechanism of the overshoot under biaxial elongational flow is different from the threading-unthreading transition under uniaxial elongational flow.

\clearpage

%

\bibliography{overshoot}

\end{document}